\documentclass[%
reprint,
superscriptaddress,
 amsmath,amssymb,
 aps,
prb,
sort&compress,numbers
]{revtex4-2}

\usepackage{float}
\usepackage{amsmath}
\usepackage{amssymb}
\usepackage{graphicx}
\usepackage[dvipsnames]{xcolor}
\usepackage{bm}
\usepackage{url}
\usepackage{pdfpages}

\renewcommand{\thetable}{\arabic{table}}  

\makeatletter
\AtBeginDocument{\let\LS@rot\@undefined}
\makeatother
\begin{document}

\title{Machine Learning Interatomic Potentials for Reactive Hydrogen Dynamics at Metal Surfaces Based on Iterative Refinement of Reaction Probabilities}

\author{Wojciech G. Stark}
\affiliation{Department of Chemistry, University of Warwick, Gibbet Hill Road, Coventry CV4 7AL, United Kingdom}
\author{Julia Westermayr}
\affiliation{Department of Chemistry, University of Warwick, Gibbet Hill Road, Coventry CV4 7AL, United Kingdom}
\affiliation{Present address: Wilhelm Ostwald Institute for Physical and Theoretical Chemistry, University of Leipzig, Leipzig 04103, Germany}
\author{Oscar A. Douglas-Gallardo}
\affiliation{Department of Chemistry, University of Warwick, Gibbet Hill Road, Coventry CV4 7AL, United Kingdom}
\affiliation{Present address: Instituto de Ciencias Qu\'imicas, Facultad de Ciencias, Universidad Austral de Chile, Isla Teja, Valdivia 5090000, Chile}
\author{James Gardner}
\affiliation{Department of Chemistry, University of Warwick, Gibbet Hill Road, Coventry CV4 7AL, United Kingdom}
\author{Scott Habershon}
\affiliation{Department of Chemistry, University of Warwick, Gibbet Hill Road, Coventry CV4 7AL, United Kingdom}
\author{Reinhard J. Maurer}%
\email{r.maurer@warwick.ac.uk}
\affiliation{Department of Chemistry, University of Warwick, Gibbet Hill Road, Coventry CV4 7AL, United Kingdom}
\affiliation{Department of Physics, University of Warwick, Gibbet Hill Road, Coventry CV4 7AL, United Kingdom}

\begin{abstract}
\section*{Abstract}
    Reactive chemistry of molecular hydrogen at surfaces, notably dissociative sticking and hydrogen evolution, plays a crucial role in energy storage and fuel cells. 
    Theoretical studies can help to decipher underlying mechanisms and reaction design, but studying dynamics at surfaces is computationally challenging due to the complex electronic structure at interfaces and the high sensitivity of dynamics to reaction barriers. In addition, \textit{ab~initio} molecular dynamics, based on density functional theory, is too computationally demanding to accurately predict reactive sticking or desorption probabilities as it requires averaging over tens of thousands of initial conditions. High-dimensional machine learning-based interatomic potentials are starting to be more commonly used in gas-surface dynamics, yet, robust approaches to generate reliable training data and to assess how model uncertainty affects the prediction of dynamic observables are not well established. Here, we employ ensemble learning to adaptively generate training data while assessing model performance with full uncertainty quantification for reaction probabilities of hydrogen scattering on different copper facets. We use this approach to investigate the performance of two message-passing neural networks, SchNet and PaiNN. Ensemble-based uncertainty quantification and iterative refinement allow us to expose the shortcomings of the invariant pairwise-distance-based feature representation in the SchNet model for gas-surface dynamics.
\end{abstract}

\maketitle

\section{Introduction} \label{sec:intro}

    Hydrogen evolution reaction on metal surfaces is a process with a great variety of practical applications, including energy storage, fuel cells~\cite{kirchheim_hydrogen_1988}, or corrosion of spent nuclear fuel~\cite{liu_roles_2016}. Reactive hydrogen chemistry at surfaces plays an important role in key heterogeneous catalysis processes~\cite{sabbe_first-principles_2012} such as the Haber-Bosch process~\cite{ertl_reactions_2008, smith_current_2020} or methanol synthesis~\cite{waugh_methanol_1992,grabow_mechanism_2011,behrens_active_2012}. Copper is a particularly interesting metal to study in this context due to its widespread use as a catalyst in many chemical reactions. Consequently, the dissociation of hydrogen on Cu surfaces has become a benchmark problem for experimental~\cite{anger_adsorption_1989,berger_search_1990,berger_investigation_1991,rettner_quantumstatespecific_1995,hodgson_rotational_1997,hou_stereodynamics_1997,murphy_adsorption_1998,kaufmann_associative_2018,chadwick_stopping_2022} and theoretical~\cite{luntz_how_2005,salin_theoretical_2006,diaz_chemically_2009,diaz_dynamics_2010,marashdeh_surface_2013,mondal_thermal_2013,nattino_dissociation_2014,jiang_six-dimensional_2014,cao_hydrogen_2018,spiering_testing_2018,smeets_quantum_2019,zhu_unified_2020,galparsoro_first_2020,dutta_effect_2021,smits_accurate_2022,smits_quantum_2022,zhu_investigating_2023} research.

    \textit{Ab~initio} molecular dynamics (AIMD) simulations utilize electronic structure methods, typically through density functional theory (DFT). Simulating molecules using on-the-fly AIMD methods is typically feasible only for short-time scales, usually in the range of a few picoseconds, small systems, in the range of a few dozen atoms at most, and few trajectories. Furthermore, simulating the reactive dynamics of molecules at metal surfaces is exceptionally challenging due to the complex electronic structure of metallic surfaces and the high dimensionality of the system involving adsorbate and substrate dynamics. Most quantities of interest in gas-surface dynamics, such as the probability of dissociative sticking or recombinative desorption, require a high sensitivity towards barriers and energy landscape corrugation. Finally, to compare theoretical calculations with experimental results, ensemble averaging over large numbers of trajectories is often necessary. This makes AIMD unfeasible if converged macroscopic experimental observables are to be simulated. Therefore, to satisfy these requirements, analytical surrogate models of the potential energy surfaces (PESs) are typically employed. Various techniques were used in the past to model the dissociative chemisorption of H\textsubscript{2} at metal surfaces~\cite{kroes_quantum_2016,kroes_computational_2021}, such as potentials based on the corrugation-reducing procedure (CRP)~\cite{busnengo_representation_2000,busnengo_surface_2005} for Cu(111)~\cite{salin_theoretical_2006,diaz_chemically_2009,diaz_dynamics_2010,smeets_quantum_2019,smeets_designing_2021, smits_accurate_2022}, Cu(211)~\cite{smeets_quantum_2019}, Cu(100)~\cite{marashdeh_surface_2013}, Pd(100)~\cite{lozano_adsorption_2009} and Ni(111)~\cite{tchakoua_toward_2019}, its extension, dynamic corrugation model (DCM)~\cite{smits_beyond_2021} for Cu(111)~\cite{smits_quantum_2022,smits_quantum_2023}, the modified Shepard interpolation~\cite{ischtwan_molecular_1994,thompson_molecular_1997} for Cu(111)~\cite{diaz_dynamics_2010} and Pt(111)~\cite{crespos_multi-dimensional_2003,crespos_application_2004}, or the permutation invariant polynomials (PIPs)~\cite{braams_permutationally_2009,bowman_high-dimensional_2011} employing neural networks (PIP-NN)~\cite{jiang_permutation_2013} for Cu(111), Ag(111)~\cite{jiang_six-dimensional_2014,jiang_permutation_2014} and Co(0001)~\cite{hu_site-specific_2015,jiang_six-dimensional_2015}. Besides the PIP-NN, other machine-learning (ML)-based models were employed to study hydrogen chemistry at metal surfaces in the past, such as the embedded atom neural network (EANN)~\cite{zhang_embedded_2019}, which was also successfully employed for modeling dissociative chemisorption at multiple Cu surfaces~\cite{zhu_unified_2020}.
    Although these methods have proven to be robust and accurate, most of them have clear limitations. For example, the CRP has not yet been applied beyond the gas-surface dynamics of homonuclear diatomic molecules. Additionally, CRP models have reduced dimensionality, typically including only the degrees of freedom of the molecule on frozen surfaces. While approximate approaches to include phonon and temperature effects have been proposed~\cite{bonfanti_7d_2013,marashdeh_surface_2013,kroes_vibrational_2017,dutta_effect_2021,smits_accurate_2022,smits_quantum_2022}, the accuracy and efficiency of these approximations have to be verified for each system. The modified Shepard interpolation method was found to have symmetry-related issues, which have been addressed in a H\textsubscript{2}/Pd(111) study~\cite{abufager_modified_2007}. The potentials based on this method are restricted to low-dimensional systems and exclude surface degrees of freedom. PIP-NN is capable of representing high-dimensional systems accurately, however, this comes with unfavorable computational scaling with the number of atoms in the system, due to a large number of required PIPs. Even with recent improvements~\cite{shao_communication_2016}, PIP-NNs cannot be easily extended to consider all degrees of freedom for extended surfaces. 
    
    Recently, for modeling PESs, attention has shifted to ML-based interatomic potentials (MLIPs), due to their simplicity, flexibility, accuracy, and ability to model all degrees of freedom~\cite{behler_four_2021, deringer_machine_2019, kocer_neural_2022}. Many approaches to MLIPs have appeared in recent years, with neural networks (NNs) being one of the most prominent. Behler and Parrinello introduced high dimensional NNs~\cite{behler_generalized_2007}, an atomistic approach to NNs, in which atomic energies are learned, to be finally summed up to approximate the PES of the system. One of the biggest challenges in creating MLIPs is the representation of atomic environments in the system, which has been the subject of extensive studies within the ML community in recent years~\cite{musil_physics-inspired_2021,zhang_atomistic_2022,langer_representations_2022}. Atomic feature representations should capture the dependence of the atomic energy with respect to the environment. They typically achieve this by expanding the atomic neighborhood into a functional dependence of two-body (distances), three-body (angles), and higher-order terms. Furthermore, atomic features should ensure invariance of the PES with respect to geometric symmetries, such as rigid rotation, inversion, and translation -- symmetry operations that comprise the Euclidean group $E(3)$. The most commonly employed symmetry-invariant descriptors include atom-centered symmetry functions (ACSFs)~\cite{behler_atom-centered_2011} and smooth overlap of atomic positions (SOAP)~\cite{bartok_representing_2013}, but many other approaches have been developed~\cite{huo_unified_2022,rupp_fast_2012}. 
    Recently, the importance of symmetry equivariance in atom-centered features was concurrently acknowledged by several research groups. This means that descriptors are able to describe the transformation properties of vectorial and tensorial features (e.g. atomic force vectors) with respect to rigid rotations and symmetry transformations in the atomic environment. Together with the emergence of message-passing neural networks (MPNNs), which consider molecules and materials as graphs and directly learn atom-centered descriptors,~\cite{gilmer_neural_2017,schutt_quantum-chemical_2017} this has given rise to several new deep-learning-based MLIP architectures~\cite{schutt_equivariant_2021,batzner_e3-equivariant_2022,unke_spookynet_2021,musaelian_learning_2023,qiao_informing_2022,haghighatlari_newtonnet_2022,batatia_mace_2022}. These new approaches provide a more complete representation of the atomic environments with respect to geometric changes and have shown to be more data efficient and accurate in learning MLIPs based on consistent sets of energies and forces for a variety of molecular and bulk materials systems~\cite{schutt_equivariant_2021,batzner_e3-equivariant_2022,batatia_mace_2022}. A commonly employed MLIP architecture in gas-surface dynamics is EANN~\cite{zhang_embedded_2019}, which employs 2-body and 3-body terms to represent the atomic environment but currently does not implement feature equivariance. Nevertheless, it is able to achieve highly accurate and efficient model representations. The relevance of feature equivariance for gas-surface dynamics simulations has not yet been assessed in detail.
    One of the biggest difficulties in developing MLIPs is the creation of a structurally diverse, meaningful, yet efficient, and sparse data set that will enable the creation of reliable models. Adaptive sampling and active learning techniques~\cite{behler_constructing_2015,botu_adaptive_2015,li_molecular_2015,gastegger_machine_2017,podryabinkin_active_2017,smith_less_2018,van_der_oord_hyperactive_2022,kulichenko_uncertainty-driven_2023}, based on an iterative search for high-error structures, has become a state of the art method in computational materials research. Despite the success of these techniques, model errors are typically measured on test data sets. Performance measures focused on dynamic observables and simulation results provide for more reliable validation.

    The probability of dissociative sticking of hydrogen or recombinative desorption crucially depends on the barrier for dissociation or recombination, respectively. In addition to the necessity of accurate and robust MLIPs, identifying a reliable approximation to DFT can be challenging. Initially, the works that investigated hydrogen dynamics at Cu surfaces employed generalized gradient approximation functionals such as PW91 or RPBE~\cite{luntz_how_2005,salin_theoretical_2006,diaz_dynamics_2010}. However, it was observed that RPBE underestimates and PW91 overestimates reaction probabilities for H\textsubscript{2} dissociation at Cu(111) surfaces~\cite{diaz_chemically_2009}. In order to address this issue for hydrogen surface chemistry, the specific reaction parameter (SRP) functional~\cite{gonzalez-lafont_direct_1991} was developed and applied in many studies of dissociation at the Cu(111) surface~\cite{diaz_chemically_2009,kroes_vibrational_2017,cao_hydrogen_2018,spiering_testing_2018,smits_accurate_2022,smits_quantum_2022,smits_quantum_2023}. The SRP48 functional employed in most of those studies, has been additionally proven to be transferable to other surfaces beyond Cu(111), such as Cu(100)~\cite{sementa_reactive_2013,marashdeh_surface_2013} or Ag(111)~\cite{nour_ghassemi_test_2018}. It was also successfully applied to stepped surfaces~\cite{migliorini_surface_2017}, including H\textsubscript{2}/Cu(211) systems~\cite{cao_hydrogen_2018,fuchsel_anomalous_2018,smeets_quantum_2019,dutta_effect_2021}.
    Another, more recently introduced functional, that also shows good agreement with experiments, is optPBE-vdW~\cite{klimes_chemical_2010}. This functional additionally includes long-range van der Waals interactions and was employed for H\textsubscript{2}/Cu(111) and Cu(100) systems where it was shown to give predictions for reactive sticking probabilities that are comparable to SRP48.~\cite{wijzenbroek_performance_2015}. The optPBE-vdW functional was also utilized in the unified EANN-based model for H\textsubscript{2} dynamics at multiple facets of copper recently developed by Zhu~\textit{et~al.}, giving an excellent agreement with experiments for all of the modeled surfaces~\cite{zhu_unified_2020}. In this work, we will use the SRP48 functional due to its prior successful application for various properties beyond sticking probabilities \cite{fuchsel_anomalous_2018, spiering_testing_2018, smeets_quantum_2019, smits_quantum_2023}.

    In this paper, we perform adaptive data generation with uncertainty quantification (UQ) to construct MLIPs based on message-passing for the simulation of reactive scattering of molecular hydrogen on copper. Because of the impact the surface termination of the catalyst has on the chemical reactivity, we include multiple copper facets in our model. By directly targeting dynamic observables and by providing UQ throughout the training process, we are able to assess the accuracy and data efficiency of MLIPs with or without equivariant features for two similar architectures, namely the SchNet model~\cite{schutt_quantum-chemical_2017,schutt_schnet_2018,schutt_schnetpack_2019} and its equivariant successor, the polarizable atom interaction NN (PaiNN)~\cite{schutt_equivariant_2021}. We present our workflow for adaptive sampling driven by the direct simulation of sticking probabilities at various incidence energies and vibrational initial states of the molecule. While the invariant SchNet model requires many adaptive sampling iterations to converge to a smooth and accurate PES, PaiNN achieves similar or better accuracy already after one adaptive sampling iteration, providing evidence of the shortcomings of the SchNet model for this application area. SchNet-based MLIPs converge only very slowly to smooth PESs that yield reaction probabilities that agree with reference results, whereas PaiNN provides robust and accurate simulated results with only a fraction of the training data. We can trace these results back to the difficulty of constructing smooth PESs with SchNet, which directly affects the simulated reaction probabilities. By employing larger unit cells than in previously reported models for this system, we are able to describe the low coverage limit and enable a straightforward extension to study phenomena beyond sticking, such as recombinative desorption.

\section{Methods} \label{sec:methods}

    \subsection{Message-passing-NN-based interatomic potentials} \label{sec:methods_ips}
    
        In message-passing NNs (MPNNs), each atom (node) is connected with edges to all neighboring atoms within a specified cutoff distance, to create graphs embedded in 3-dimensional Euclidean space. The MPNN graph layout enables the passing of information between atoms through a series of message passing-update steps, creating a representation that indirectly carries details about atoms from outside of the cutoff distance and has the ability to encode many-body interactions. This process can be described as
        \begin{equation} \label{eqn:message}
            \textbf{m}^{t+1}_{i} = \sum_{j\in \mathcal{N}(i)} \textbf{M}_{t}(\textbf{s}^{t}_{i},\textbf{s}^{t}_{j},\vec{e}_{ij}),
        \end{equation}
        \begin{equation} \label{eqn:h_eq}
            \textbf{s}^{t+1}_{i} = \textbf{U}_{t}(\textbf{s}^{t}_{i},\textbf{m}^{t+1}_{i}),
        \end{equation}
        where $\textbf{m}^{t+1}_{i}$ is a message created by summing over the scalar atomic features $\textbf{s}$ of nodes (atoms) $i,j$ at step $t$, connected by edge features $\vec{e}$ (e.g. relative positions of nodes $\vec{e}_{ij}=\vec{e}_{j}-\vec{e}_{i}$). $\textbf{M}_{t}$ and $\textbf{U}_{t}$ are nonlinear message and update functions, respectively~\cite{gilmer_neural_2017}.
        
        Recently, equivariant (vectorial) features embedded in MPNNs proved to provide significant improvement in the data efficiency and accuracy of the models~\cite{schutt_equivariant_2021, batzner_e3-equivariant_2022,musaelian_learning_2023,batatia_design_2022}. To include such features in the message passing scheme, the message in Eq.~\ref{eqn:message} is adapted to
        \begin{equation} \label{eqn:message}
            \vec{\textbf{m}}^{t+1}_{i} = \sum_{j\in \mathcal{N}(i)} \vec{\textbf{M}}_{t}(\textbf{s}^{t}_{i},\textbf{s}^{t}_{j},\vec{\textbf{v}}^{t}_{i},\vec{\textbf{v}}^{t}_{j},\vec{e}_{ij}),
        \end{equation}
        where vectorial representations $\vec{\textbf{v}}$ are additionally included.
        Schütt and co-workers developed a family of MPNNs, in particular, SchNet~\cite{schutt_schnet_2018} and PaiNN~\cite{schutt_equivariant_2021}, which differ mainly by the inclusion of equivariant features in the latter. While SchNet starts from a feature embedding based on interatomic distances, PaiNN additionally propagates vectorial features. This differs from other MLIPs based on predefined distance and angle features, employing e.g. ACSF or SOAP descriptors. We are using these models here as the architectures are very similar, allowing us to study how the inclusion of equivariance affects model performance and simulation accuracy. 

    \subsection{Training data} \label{sec:methods_training_data}
    
    \begin{figure}
        \centering
        \includegraphics[width=3.4in]{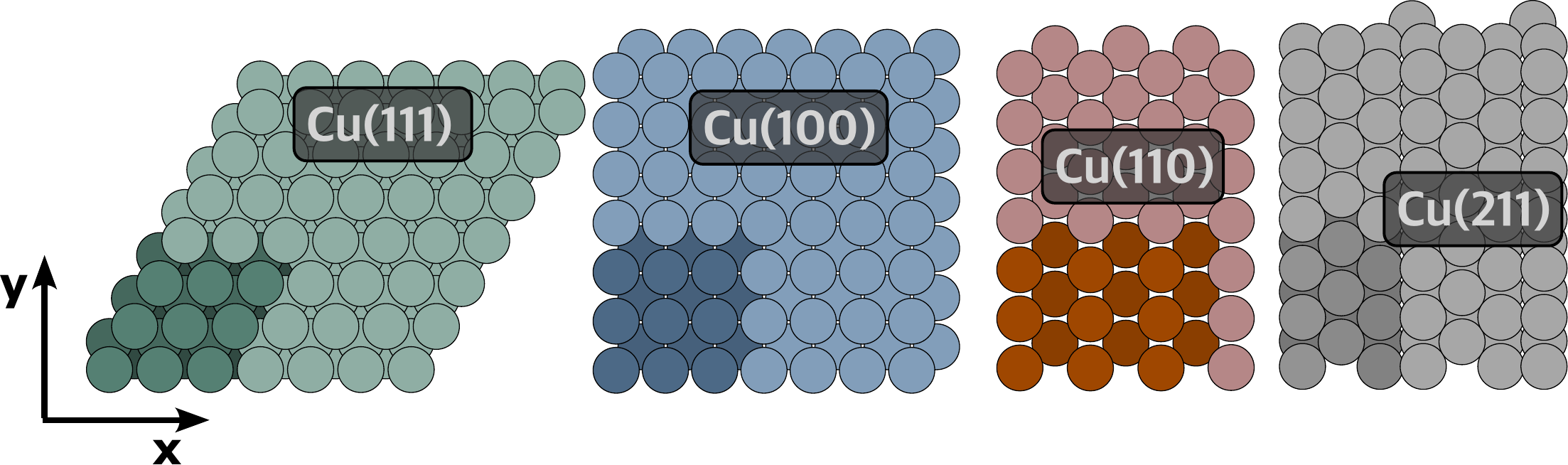}
        \caption{\textbf{Top-view of the Cu(111), Cu(100), Cu(110) and Cu(211) surfaces.} Unit cells used in the work are highlighted with darker colors.}
        \label{fig:slabs}
    \end{figure}
    
        Our MLIP is aimed at describing scattering on typical surface facets of crystalline copper. We included four Cu facets in our data set, namely Cu(111), (100), (110), and (211). In our data set we used 3$\times$3, 6-layered slabs for Cu(111), (100) and (110) surfaces and 1$\times$3, 6-layered slabs for Cu(211) surface (Fig.~\ref{fig:slabs}). Our initial data set included 2530 data points. 845 data points represent H\textsubscript{2} interacting with one of the four Cu surfaces (56 atoms) and 1685 data points represent Cu surface structures sampled at different temperatures (54 atoms). The final data set, after adaptive sampling (see below), contained 4230 data points (2545 H\textsubscript{2}/Cu and 1685 clean Cu surface structures). The final number of data points in our training data set is higher than the data set used by Zhu \textit{et al.}~\cite{zhu_unified_2020}, which was generated based on subsampling of data from many on-the-fly ab-initio MD trajectories. Our training data set generated with multiple steps of adaptive learning is densely sampled and likely could be sparsified without noticeable loss of information.
        
        The initial data set was created with data points collected from two different sources. The first source consisted of data points sampled from AIMD simulations of hydrogen atoms adsorbed at each surface at 300~K and from AIMD simulations of clean surface dynamics (without adatoms) at three different temperatures: 300, 600, and 900~K. In these calculations, only the two bottom layers of copper were fixed. The second set of data points was taken from MD simulations of H\textsubscript{2} scattering on 2$\times$2 fixed Cu surfaces, including (111), (110), and (100) facets with 4 layers. No H\textsubscript{2} scattering data was included for Cu(211) in our initial data set and thus the barrier of H\textsubscript{2}/Cu(211) dissociative adsorption is not explored directly during training of the initial models. The initial MD simulations were performed with a SchNet MLIP with standard settings trained on the data set generated and kindly provided by Jiang and co-workers~\cite{zhu_unified_2020}. The metal slabs from the data points generated in the simulations were then transcribed into 6-layered 3$\times$3 slabs and both energies and forces were calculated with DFT. We have chosen to describe the system with a minimum of 6 metal layers and larger unit cells which may be crucial to ensure convergence with respect to electron-phonon response \cite{maurer_ab_2016} as recently pointed out by Box et~al.~\cite{box_ab_2021}. Such effects may be significant for the simulation of e.g. state-to-state scattering probabilities.~\cite{spiering_testing_2018} Furthermore, including the larger unit cell will be beneficial to future-proof the data set for the study of on-surface and subsurface chemistry of hydrogen. Additionally, we added several structures with hydrogen molecules far away from the surface (around 10~$\textrm{\AA}$ above the surface), where the metal-H interaction is negligible. These structures only differ by the distance between H atoms, to ensure a robust description of the H-H bond in the gas phase. The exception was the H\textsubscript{2}/Cu(211) system, for which the initial data set contained only structures sampled from AIMD of hydrogen atoms placed on the surface (adsorbed) and several data points with a hydrogen molecule placed high above the surface with different bond lengths.

        For DFT calculations we employed the SRP exchange-correlation functional~\cite{nattino_effect_2012} containing 52\% of PBE~\cite{perdew_generalized_1996} and 48\% of RPBE functional~\cite{hammer_improved_1999} (SRP48) with a k grid of 12$\times$12$\times$1 (applies to all Cu facets used in our study). All DFT calculations were performed with the all-electron numeric atomic orbital code FHI-aims~\cite{blum_ab_2009} using a ``tight'' default basis set and the following SCF convergence criteria: tolerances of 10$^{-6}$~eV, 10$^{-3}$~eV, 10$^{-5}$~e/a$_{0}$$^{3}$ and 10$^{-4}$~eV/$\textrm{\AA}$ for the total energy, the eigenvalue energies, the charge density, and the forces, respectively.
        Minimum energy paths were obtained with climbing image nudged elastic band~\cite{henkelman_climbing_2000} calculations. We used 7, 20, and 50 images with DFT, SchNet, and PaiNN codes respectively, and maximum force along the path of 0.01 eV/\AA.

    \subsection{Adaptive sampling and uncertainty quantification strategy} \label{sec:methods_adaptive}

        \begin{figure}
            \centering
            \includegraphics[width=3.3in]{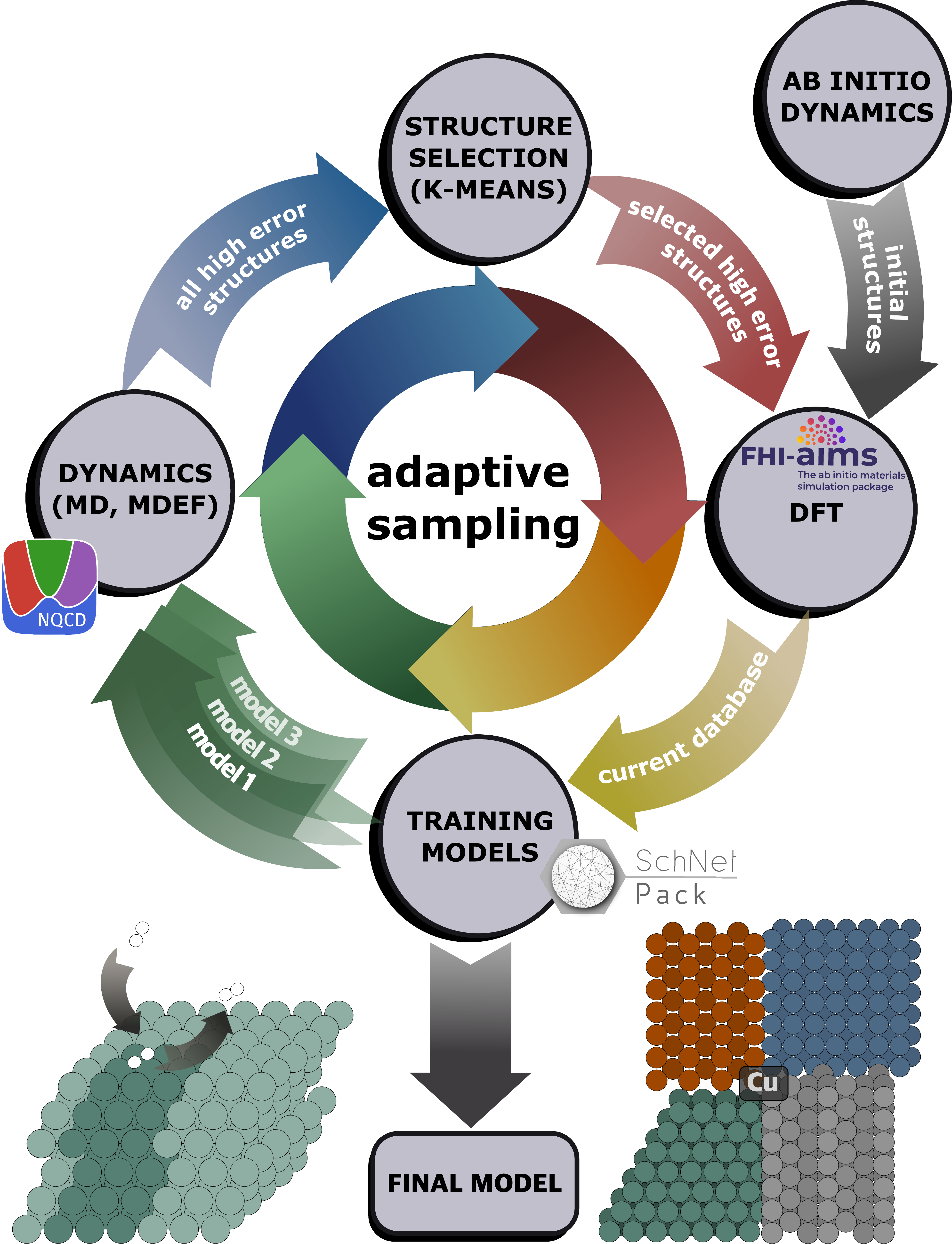}
            \caption{\textbf{Schematic representation of the iterative adaptive sampling procedure.} The initial set of DFT data points is used to train the first set of models. We directly perform gas-surface scattering dynamics to assess the validity of the potential and to select new data points for which further DFT calculations are performed and new models are trained. This process is iteratively repeated until the target dynamic observables are converged.}
            \label{fig:adaptive_learning_scheme}
        \end{figure}

        To improve the initial data set we employed an adaptive sampling procedure as shown in Fig.~\ref{fig:adaptive_learning_scheme}. This approach allows iterative improvement of the models by finding areas in phase space for which the current data points do not sufficiently constrain the model potential. To find badly represented structures, at every fifth step of the MD simulation, we evaluate the potential using 3 models trained with different train/test splits and optionally different energy/force weights for the loss function. This is a form of bootstrapping ensemble learning, which enables us to calculate the standard deviation between the potential predictions and select structures for which the standard deviation exceeds a chosen maximum error value. In such cases, models disagree and the data point is likely outside of the distribution covered by the training data. The same uncertainty estimates are later used to evaluate epistemic errors on the dynamic reaction probabilities in section \ref{sec:results}. The maximum error values were selected manually after analyzing standard deviation trends across the trajectories and their values were 0.03~eV for the first three adaptive sampling iterations and 0.025~eV for the last iterations. 
        
        For the chosen structures we calculate the associated DFT-level energies and forces. The new data points are then added to the existing data set. After having collected a full set of high-error structures for every Cu surface from reactive scattering MD simulations (see below), the data set is analyzed using k-means clustering. At every iteration of adaptive sampling, we generated between 100-200 clusters for every surface. The data points closest to the centers of the clusters were then added to a data set that contained all data points generated during this adaptive sampling step. This ensures that the smallest number of data points with the most amount of new information is added. For the k-means clustering, we used simple but informative descriptors, such as the inverse distances between hydrogen atoms and the inverse distances between every hydrogen atom and the surface, which we reduced into a 2-dimensional problem employing principal component analysis. The data set received after sampling all high-error points was then manually inspected to exclude obviously non-physical data points, which accounted for roughly 0-10\% of all sampled data points. The non-physical structures were mostly found within data points from MD simulation steps that included the Cu(110) surface. The number of such structures decreased with every iteration of adaptive sampling. In a single adaptive sampling step, the resulting data set included between 300-600 new data points depending mainly on the number of data points excluded during the last stage of data processing. The Python-based scikit-learn~\cite{pedregosa_scikit-learn_2011} package was employed for the dimensionality reduction and the clustering of the data points. All scripts and definitions related to the analysis and clustering process are provided (see Data Availability).

        For the purpose of adaptive sampling, in each iteration, we perform a full set of reactive scattering simulations for initial vibrational and rotational state distributions for states ($\mathrm{\nu}$=0, J=0) (iterations 1 to 4) and ($\mathrm{\nu}$=1, J=1) (iterations 3 to 4) of H\textsubscript{2} on all surface facets calculated at 6 different collision energies between 0.2 and 0.85~eV. The simulated sticking probabilities yield a metric for the convergence that is a dynamic observable which is directly comparable to experimental data and other computational studies. SchNet was used for all simulations regarding adaptive sampling, which ensured consistency with respect to data generation.

        Independently, we assess the performance of the MLIPs against test data with two types of errors, root-mean-square error (RMSE) and mean absolute error (MAE). Energy RMSE and MAE measures mentioned in the following sections relate to the total energy of all atoms in our system, as opposed to the errors in energy per atom.

    \subsection{Molecular dynamics simulations} \label{sec:methods_md_settings}
        
        All molecular dynamics simulations, including preparation of initial conditions, were performed using the open-source NQCDynamics.jl package~\cite{gardner_nqcdynamicsjl_2022} (\url{https://github.com/NQCD/NQCDynamics.jl}), written in the Julia programming language. AIMD simulations for the initial training data set were performed using FHI-aims~\cite{blum_ab_2009}. 
        
        To allow adaptive sampling, an interface was created within the NQCDynamics.jl code that allows calculating the standard deviation between energies obtained by separate MLIP models at every other step of the simulation (any step can be set). The package includes multiple methods and settings for creating initial conditions (e.g. Einstein-Brillouin-Keller, thermal Metropolis-Hastings Monte Carlo, and Langevin dynamics), providing an all-in-one environment for simulating dynamics in the condensed phase.

        The MD simulations of hydrogen scattering are initiated with the hydrogen molecule placed 7~$\textrm{\AA{}}$ above the surface and initial rovibrational states of H\textsubscript{2} are generated using the Einstein-Brillouin-Keller (EBK) method~\cite{larkoski_numerical_2006} for several normal incidence energies with randomly chosen polar and azimuthal angles. The maximal simulation time of every scattering trajectory was set to 3~ps with a time step of 0.1~fs. Trajectories were terminated when special conditions were met, namely when the distance between the hydrogen atoms exceeds 2.25~$\textrm{\AA}$ (trajectory counted as dissociative chemisorption event) or the average distance between the hydrogen molecule and the top surface atoms is above 7.1~$\textrm{\AA}$ (trajectory counted as scattering event). Sticking probabilities were calculated using data extracted from 10,000 trajectories for every model, surface facet, rovibrational state, and initial collision energy reported. All sticking probabilities are averaged values from the results received by 3 models with different training/test data split (in total 30,000 trajectories for every probability). All figures that report sticking probabilities also report epistemic model uncertainties as error bars, which represent the standard deviation between sticking probabilities predicted by 3 different models of the same code with different random train/test data split. Note that we do not add statistical errors as they are negligible compared to the model uncertainty. We provide evidence of this and more details on the error analysis in Supplementary Figure~S1.

        The simulations that included surfaces at 0~K were initiated using positions of the DFT-relaxed surface slabs. The simulations that included surfaces at higher temperatures were initialized with surface positions obtained from the surface-only thermal Metropolis-Hastings Monte Carlo sampling at the defined temperature and with initial hydrogen velocities based on Maxwell-Boltzmann distribution. The thermal lattice expansion was not explicitly considered.
        
        AIMD simulations of dissociated hydrogen atoms moving on all four surfaces were initiated with a hydrogen molecule adsorbed on the copper surface, relaxed at the DFT level, and the temperature of the simulation set at 300~K. Surface-only AIMD simulations were initiated using DFT-level relaxed surfaces and the simulations ran at three different temperatures (300~K, 600~K, and 900~K). Temperatures for all the AIMD simulations were controlled by the Bussi-Donadio-Parrinello thermostat~\cite{bussi_canonical_2007}. All the AIMD simulations ran for 10~ps with the time step of 5~fs.

    \subsection{Model training and optimization details} \label{sec:methods_model_training}
        
        The initial and the iteratively improved models, trained during adaptive sampling, were obtained using default SchNet settings, employing a relatively safe cutoff of 5~$\textrm{\AA}$, a batch size of 10, and varying loss function weights for energy and forces. After the fourth adaptive sampling iteration, we performed a detailed parameter optimization for both, SchNet and PaiNN models, via k-fold cross-validation (k=5), in which we divided our data set into 6 random splits of the same size. Following that, for each set of parameters, we trained 5 models, each with the same test set split, but a different validation set split. All the remaining (4) splits formed the training sets. The errors were then calculated from the predictions on validation data of the models and the best settings were chosen based on the average error of all 5 models. The final sticking probabilities were calculated employing optimized models. As energies and forces were trained together, these terms entered the same loss function. For all the final models, we used loss function energy and force weights of 0.05 and 0.95 respectively, however sticking probabilities shown in Fig.~\ref{fig:sticking_cu111_painn} were calculated using models with 0.5 and 0.5 weights for energy and force, which we used initially for PaiNN models (default settings in PaiNN). However, for an even better comparison, we continued with the 0.05 energy and 0.95 force weights for all the models. The difference in energy errors for PaiNN models with both weights was negligible, although, the force errors were roughly 2 times lower for the 0.05 energy and 0.95 force weights. Other settings are provided in section~\ref{sec:optimization}.
        SchNet was accessed via SchNetPack v1.0.0~\cite{schutt_schnetpack_2019} (\url{https://github.com/atomistic-machine-learning/schnetpack}) master branch and PaiNN through the same repository, but through the developmental ("dev") branch, that is currently a base for the SchNetPack v2.0~\cite{schutt_schnetpack_2023}.

\section{Results}\label{sec:results}

    \subsection{Model parameter optimization and learning behavior} \label{sec:optimization}
    
        We performed an optimization of different model parameters for both SchNet and PaiNN models independently, using k-fold cross-validation (k=5). Supplementary Figure~S2 shows the convergence of the number of features and interaction blocks (left) and cutoff distance (right) for both models with respect to force RMSEs in log scale. After analyzing the results of the optimization, we concluded that we will use 7 interactions, 512 features, and a cutoff distance of 4~$\textrm{\AA}$ in the final models for both codes. More information about the model parameter optimization can be found in the SI.

        \begin{figure}
            \centering
            \includegraphics[width=3.5in]{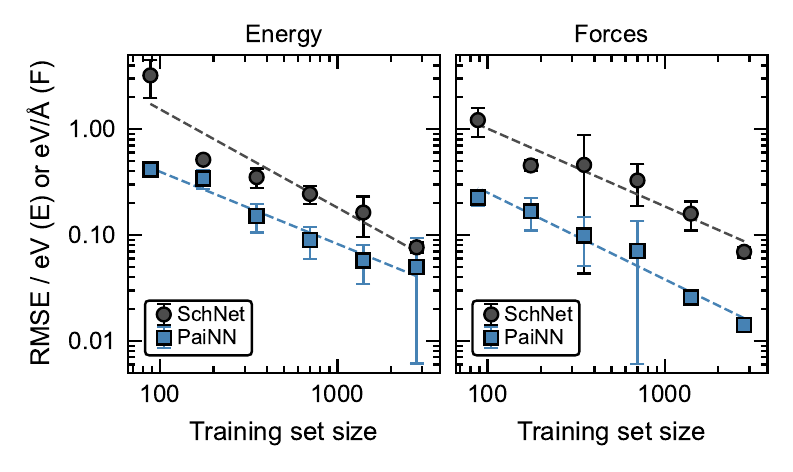}
            \caption{\textbf{Learning curves for SchNet and PaiNN models.} Log-log plot of the test set predictive error (RMSE) dependence on the size of the training set for both models (SchNet and PaiNN) with respect to energy (left) and force (right) values. The error bars represent the standard deviation between the RMSEs obtained for all splits.}
            \label{fig:learning_curves}
        \end{figure}

        To ensure proper model training, we have generated learning curves (shown in Fig.~\ref{fig:learning_curves}) based on energy and force RMSE convergence with respect to the training set size. Overall, the learning rates for SchNet and PaiNN are comparable for energies and forces, but the curves generated with PaiNN models are shifted, providing prediction errors 2-5 times lower across all training set sizes. This shift associated with the inclusion of equivariant features was also reported by Batzner~\textit{et~al.}~\cite{batzner_e3-equivariant_2022} and Batatia~\textit{et~al.}~\cite{batatia_mace_2022}. Batzner~\textit{et~al.} additionally observed a change in the shape of the force learning curve, caused by equivariant features. We do not observe this in our data. The energy learning rate obtained with SchNet models is slightly higher than with the equivariant PaiNN models, although just by looking at the uncertainties, we can assume that the difference may not be statistically significant. Even though the energy learning rate for SchNet might be slightly higher, the energy RMSEs of PaiNN are still lower for any training set sizes. We can conclude that the PaiNN models based on equivariant features are superior in representing the atomic environment, in particular for predicting equivariant quantities such as forces. The learning curves indicate that such models require considerably less data to achieve the same model errors than comparable invariant MPNN models.

    \subsection{Adaptive sampling of reaction probabilities} \label{sec:results_adaptive_sampl}

        \begin{figure*}
            \centering
            \includegraphics[width=\linewidth]{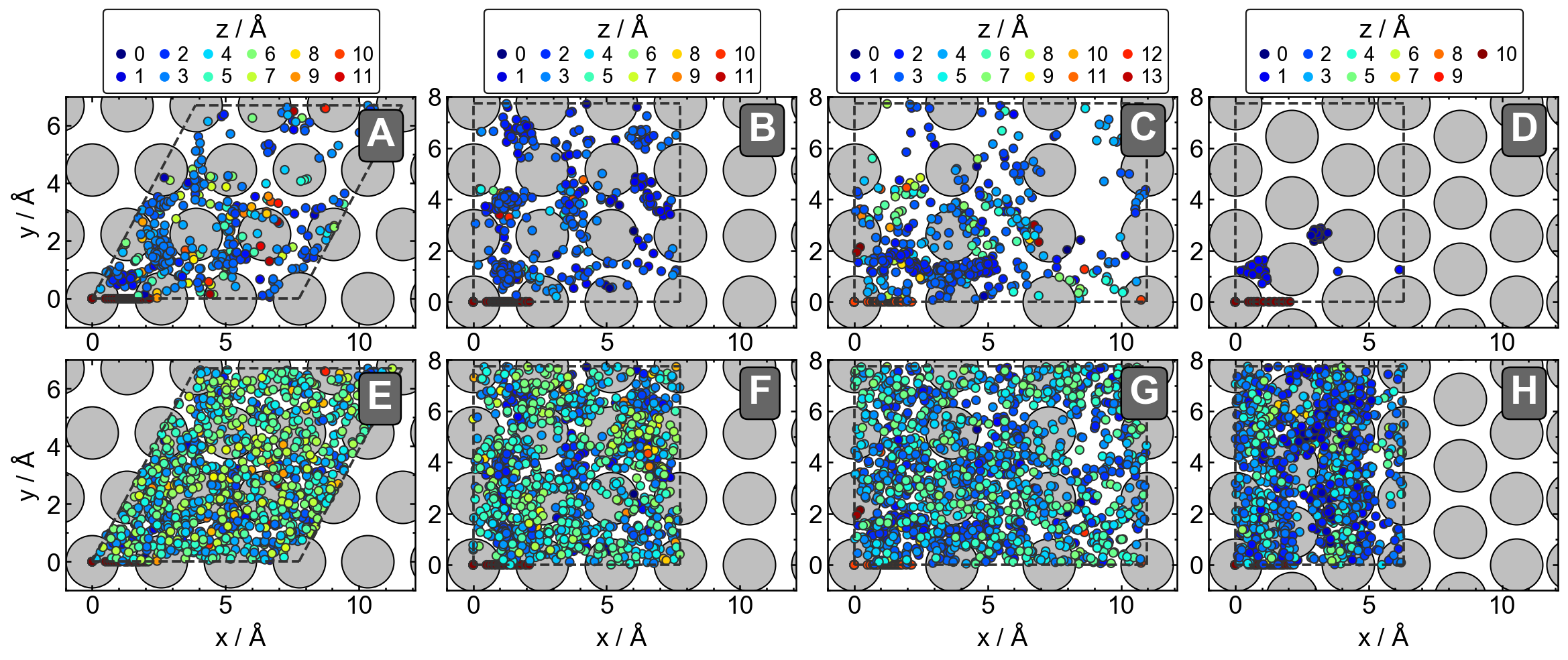}
            \caption{\textbf{Distribution of the hydrogen atoms within the unit cell.} Panels show x and y coordinates of the unit cells of all structures with colored circles depicting the H atom positions in the initial (top) and the final adaptively improved (bottom) training data sets for all 4 surfaces: Cu(111) (A, E), Cu(100) (B, F), Cu(110) (C, G), Cu(211) (D, H). The circle color indicates the value of height above the top layer of the respective Cu surface. Top surface atoms are schematically represented as large grey circles. As the positions of surface atoms are different for most of the structures, the surface atom positions are shown for the DFT-level relaxed surfaces.}
            \label{fig:adaptive_sampl_h2oncu_xyz}
        \end{figure*}
        
        To improve the ability of the trained MLIPs to describe the reactive scattering of H\textsubscript{2}, we carried out adaptive sampling according to the scheme in Fig.~\ref{fig:adaptive_learning_scheme}. The adaptive sampling loop was iterated 4 times until we obtained a satisfactory level of accuracy of the desired output property, namely the sticking probability as a function of incidence energy and the initial vibrational and rotational state of the molecule. Fig.~\ref{fig:adaptive_sampl_h2oncu_xyz} shows the distribution of all hydrogen atom positions from all data points contained in the initial and final training data set (before and after adaptive sampling). The hydrogen atom distribution in the final data set is noticeably more diverse, with hydrogen molecules exploring the entire unit cell. Supplementary Figure~S3 furthermore shows that the different adaptive sampling steps add data points across a wide range of H-Cu and H-H distances throughout the entrance channel and dissociation barrier region for all surfaces.

        \begin{figure}
            \centering
            \includegraphics[width=3.4in]{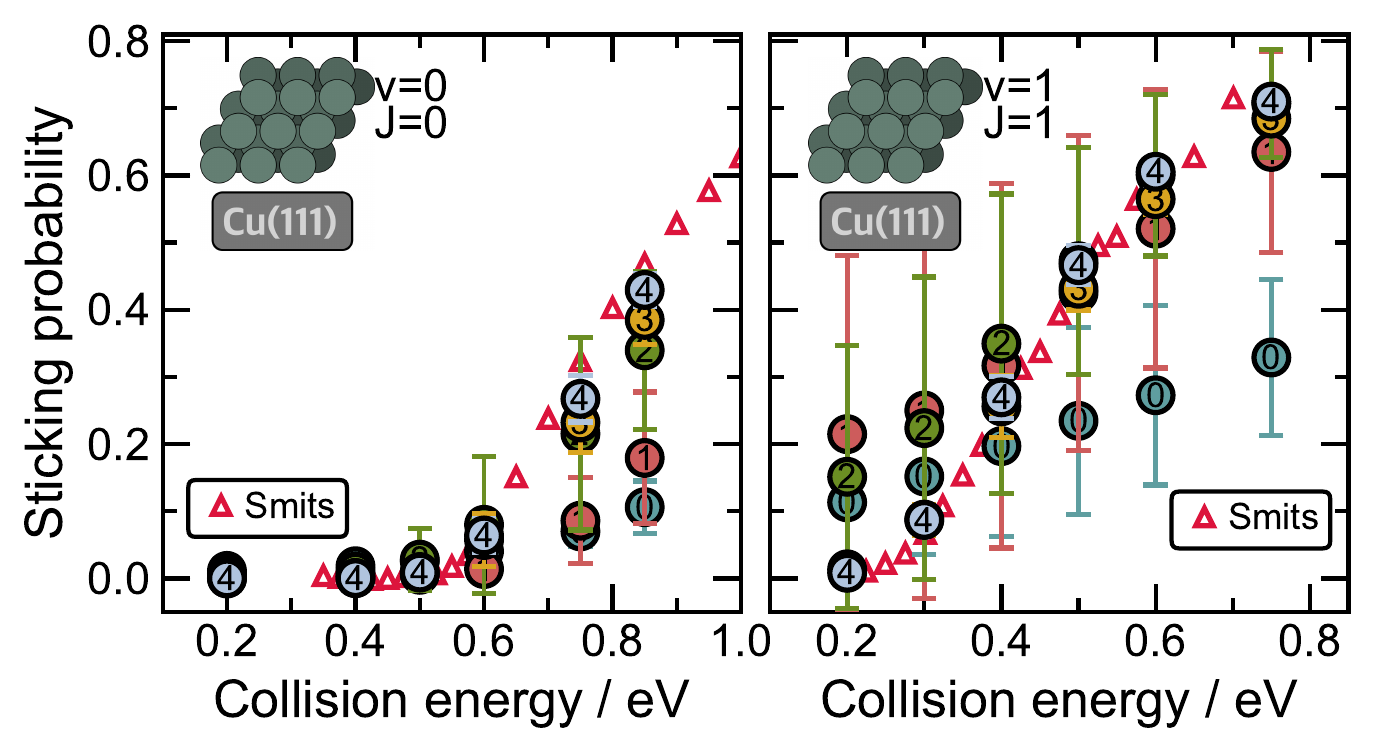}
            \caption{\textbf{SchNet predicted sticking probabilities.} Sticking probabilities have been calculated for all adaptive sampling iterations at different collision energies for H\textsubscript{2} scattering on Cu(111) at 0~K with molecules prepared in the ground state, ($\mathrm{\nu}$=0, J=0) (left), and the rovibrationally excited state, ($\mathrm{\nu}$=1, J=1) (right). Each point represents an averaged value of sticking probability predicted using a committee of 3 models based on different train/test splits. For each initial condition, 10,000 trajectories were used for ensemble averaging. Colored circles represent results calculated for different adaptive sampling iterations. Numbers inside the circles refer to the respective iteration, with "0" representing the initial model. Both figures include QCD-BOSS reference data, depicted as red triangles (\textbf{{\color{red}$\pmb{\triangle}$}}), reported by Smits~\textit{et~al.}~\cite{smits_quantum_2022}}.
            \label{fig:sticking_cu111_schnet}
        \end{figure}
        
        Fig.~\ref{fig:sticking_cu111_schnet} presents the simulated sticking probabilities for the reactive scattering of H\textsubscript{2} molecules on a 0~K Cu(111) surface in the vibrational ground and first excited state based on different iterations of the SchNet MLIP. Error bars correspond to model uncertainties calculated with ensemble learning. We compare our simulation results to the literature data based on quasi-classical dynamics (QCD) simulations with a CRP potential, reported by Smits~\textit{et~al.}~\cite{smits_quantum_2022}. These results are based on the Born-Oppenheimer static surface (BOSS) approximation, which neglects the movement of the metal surface atoms and employs the same (SRP48) functional as we use in this study for DFT calculations. We will address this model as QCD-BOSS. In order to compare our results meaningfully to the literature references, we simulate scattering at initially cold metal surfaces.
        
        The initial training data set marked with 0 in~Fig.~\ref{fig:sticking_cu111_schnet} is clearly not sufficient to capture the relevant barriers and features of the energy landscape to describe sticking on Cu(111), neither in terms of absolute prediction nor in terms of standard deviations that arise from the model uncertainty. The barrier for H\textsubscript{2} dissociation on Cu(111) as predicted by the SRP48 functional is 0.636~eV (bridge site)~\cite{smeets_designing_2021}. For collision energies below this value, the initial MLIP correctly predicts vanishing ($\mathrm{\nu}$=0, J=0) sticking probability, but for energies above the barrier, sticking is heavily underestimated. At the same time, the probability of dissociative chemisorption at low collision energy for the ($\mathrm{\nu}$=1, J=1) state is overestimated compared to literature data. The former suggests that the barrier of the MLIP is too high, and the latter suggests that the shape of the entrance channel to the barrier is incorrect, promoting the vibrational enhancement of sticking.
        
        In the first two iterations of adaptive sampling, we exclusively included the H\textsubscript{2} ($\mathrm{\nu}$=0, J=0) sticking probability simulations. Already at the second iteration, the sticking probabilities are within 20\% of the literature results for ($\mathrm{\nu}$=0, J=0) state across all collision energies, whereas H\textsubscript{2} ($\mathrm{\nu}$=1, J=1) scattering in the second iteration still yields too much sticking at low collision energies. The third and fourth iterations included data sampled from the H\textsubscript{2} ($\mathrm{\nu}$=1, J=1) runs, which then leads to a rapid convergence to the correct description of the entrance channel and virtually no sticking at low collision energies from the third iteration onwards. For both the ground and excited states, the third and fourth iterations provide qualitatively correct sticking probability curves, whereas the fourth iteration provided a further numerical refinement that brings the results into satisfactory agreement with the reference data. 
        
        In addition to the iterative improvement of the predictions in each iteration, we further find that the standard deviations calculated via the ensemble of models are considerably reduced in the third and fourth iterations. This is an indication that the models confidently and consistently predict the relevant phase space as any random 80:10 train/test split of the data (10\% held for validation) leads to an accurate description of the energy landscape.
        
    \subsection{The role of equivariant features}   \label{sec:results_equiv_vs_non}

        In the previous section, the convergence of the sticking probability during adaptive sampling was discussed for the SchNet model. We showed that four iterations of the adaptive sampling procedure were required to achieve a well-balanced and representative data set to create an accurate and robust MLIP for gas-surface scattering. We now investigate how the equivariant model PaiNN performs based on the exact same data. For this, we take the same series of initial and adaptively improved data sets and train a PaiNN model for each iteration. In doing so, we rely on the phase space explored by the adaptive sampling that was performed with SchNet.
        
        \begin{figure}
            \centering
            \includegraphics[width=2.3in]{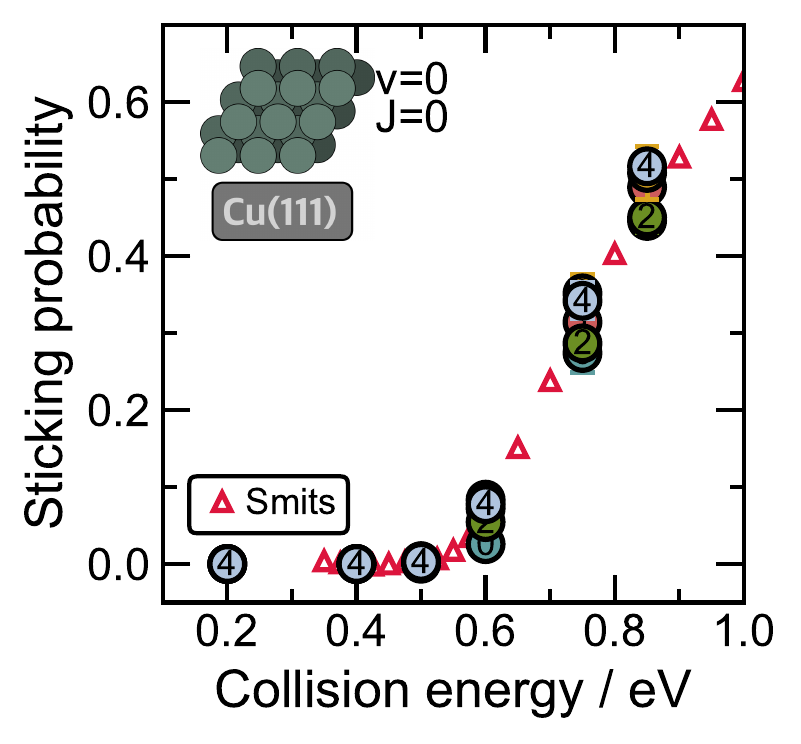}
            \caption{\textbf{PaiNN predicted sticking probabilities.} Sticking probabilities for all the adaptive sampling iterations at different collision energies for H\textsubscript{2} scattering on Cu(111) at 0~K with molecules prepared in the ground state, ($\mathrm{\nu}$=0, J=0). Each point represents an averaged value of sticking probability predicted using a committee of 3 models based on different train/test splits. For each initial condition, 10,000 trajectories were used for ensemble averaging. Colored circles represent results calculated for different adaptive sampling iterations. Numbers inside the circles refer to the respective iteration, with "0" representing the initial model. QCD-BOSS reference data reported by Smits~\textit{et~al.}~\cite{smits_quantum_2022} is depicted as red triangles (\textbf{{\color{red}$\pmb{\triangle}$}}).
            \label{fig:sticking_cu111_painn}}
        \end{figure}

        As shown in Fig.~\ref{fig:sticking_cu111_painn}, even for the initial training data set, the sticking probabilities predicted with the PaiNN model closely match the literature reference values. After the first iteration, we achieve almost perfect agreement with the reference probabilities and further iterations only provide minimal refinement of the dynamically averaged simulation results. This suggests that the failure of the initial SchNet model in the early iterations was not based on insufficient coverage of the phase space by the training data, but rather by the inability of SchNet to learn a reliable MLIP based on this data set. The results confirm our conclusions based on the learning curves (see Fig.~\ref{fig:learning_curves}) that the introduction of equivariant features provides for much more efficient learning based on sparse data. If we had used the equivariant PaiNN model for the adaptive sampling, we would likely have reached a converged model with only one or two iterations and substantially fewer data points. This is a crucial advantage as the evaluation of DFT reference data comes with a considerable computational cost for periodic slab systems.

        \begin{figure*}
            \centering
            \includegraphics[width=\linewidth]{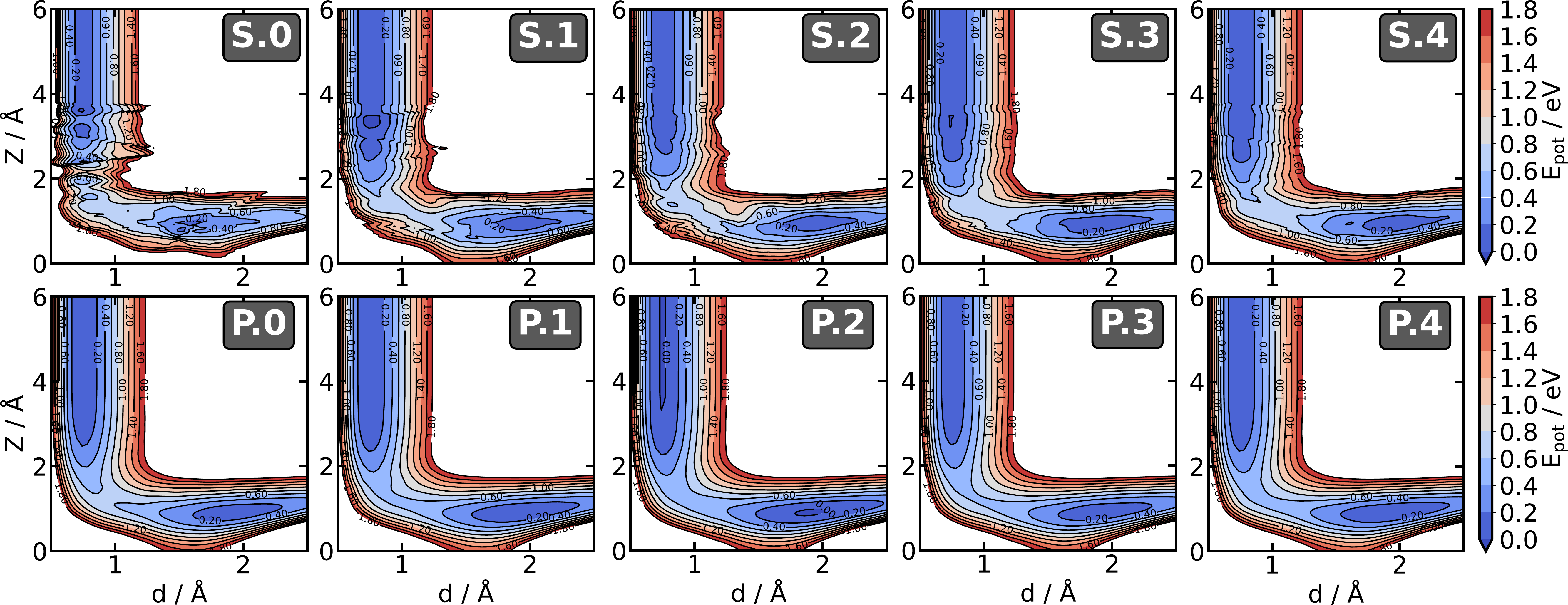}
            \caption{\textbf{Elbow plots (2d cuts through the PES) generated with different iterations of the SchNet (\textit{S.0-4}) and PaiNN (\textit{P.0-4}) models.} The initial model and adaptive sampling iterations are labeled with \textit{0} and \textit{1-4}, respectively. The elbow plots were constructed by taking the transition state structure for dissociative chemisorption of H\textsubscript{2} on Cu(111) and varying the center of mass height (Z) and the molecular interatomic distance (d).}
            \label{fig:elbow_init_last}
        \end{figure*}
        
        To obtain further insight into the origin of the failure of SchNet to capture the correct gas-surface dynamics, we analyzed cuts through the PESs. The traditional depiction to study the relevant degrees of freedom during the reactive scattering of diatomic molecules on surfaces is the ``elbow'' plot, which depicts the PES contour as a function of the center of mass distance between molecule and surface and the hydrogen-hydrogen interatomic distance. A reactive pathway is depicted as a vertical approach to the transition state, followed by a dissociation of the bond. Fig.~\ref{fig:elbow_init_last} depicts elbow plots for all iterations of the SchNet and PaiNN models for H\textsubscript{2} on Cu(111). The PES obtained with the initial SchNet model shows significant artifacts and discontinuities. Considering the model irregularities, it is surprising that we find relatively few instances of model failure during the dynamics. Successive iterations of increasing the training data set with relevant information remove these artifacts, but even in the third and fourth iterations, the PES as predicted by the SchNet model is far from smooth, which is surprising and worrying.

        In contrast, the PaiNN model, which directly trains vector-valued features to represent forces in the training data, yields smooth and well-behaved PESs already with the initial training data. We can see that successive iterations of the training data set mostly lead to a refinement of the model around the barrier region. 
        
        The issues with the SchNet energy landscapes become more evident when studying the relevant barriers along the minimum energy paths of dissociative chemisorption for SchNet and PaiNN. Supplementary Figure~S4 shows the minimum energy paths for the four surface facets predicted by SchNet, PaiNN, and the DFT reference. In all cases, the energy barriers as predicted by SchNet are not smooth and due to the presence of artificial local minima converge slowly. However, the final reaction barrier values are not far from the reference results for most surfaces, which explains the acceptable predictions of sticking probabilities, at least at 0~K. In the case of Cu(110), the barrier predicted by the SchNet model is significantly underestimated and the shape of the energy landscape is different than the shapes of the barrier generated with the PaiNN model and DFT. On the other hand, the barriers predicted by the PaiNN model are smooth and match the entire minimum energy paths predicted by the DFT very well for all the surface facets.

    \subsection{Model error analysis} \label{sec:results_errors}

        The learning curves (Fig.~\ref{fig:learning_curves}) imply that equivariant PaiNN models enable much lower force errors and considerably lower energy errors than SchNet models based on the same training data. When we compare the energy and force predictions of the best final SchNet and PaiNN models against the DFT-based reference values (see Supplementary Figure~S5), we find that both models provide accurate energy predictions across all four studied surface facets with few outliers. However, it is clear that the energy predictions by the PaiNN model are much more robust and consistent. The PaiNN model provides a three times lower MAE and a more than four times lower RMSE. It is clear that total energy predictions with SchNet are relatively accurate, but SchNet models do not necessarily provide smooth energy landscapes. Note that the reported energies are total energies over the whole system, so they may not be good indications of the errors associated with changes in relative energy due to the motion of hydrogen with respect to the copper atoms.
        
        The inability of SchNet to describe smooth energy landscapes for this data set becomes evident when studying the predicted force errors of SchNet and PaiNN (Supplementary Figure~S5, panels S.c and P.c) as forces better capture the model accuracy for individual atomic motion. The SchNet force predictions show visible outliers across all surface facets, whereas the PaiNN predictions are in excellent agreement with respect to the reference values for both energies and forces. Both force MAEs and RMSEs received for the best final models are more than 5 times lower for equivariant PaiNN models than for the SchNet models, which is in line with the learning curve errors and rapid adaptive sampling convergence. While PaiNN improves upon SchNet at predicting energies, it is even better at predicting forces. By providing consistently accurate energy and force prediction, PaiNN has the ability to yield much smoother PES.
        
        \begin{table}
           \caption{\textbf{Averaged test RMSEs and MAEs for initial and final SchNet and PaiNN models.} Predicted errors of energies (meV) and forces (meV/$\textrm{\AA}$) using initial and final SchNet and PaiNN models. Errors are listed for all systems and are additionally broken down into the four studied surfaces: Cu(111), (100), (110), and (211). All the structures that include H\textsubscript{2} contain 56 atoms. Clean surface structures contain only 54 Cu atoms. The errors are averaged over 5 models based on different random train/test sets.}
           \begin{tabular}{l|cccc|cccc} \hline \hline
               \multicolumn{1}{c}{} & \multicolumn{4}{c}{SchNet} & \multicolumn{4}{c}{PaiNN} \\ 
               \multicolumn{1}{c}{ } & \multicolumn{2}{c}{Energy}  & \multicolumn{2}{c}{Forces} & \multicolumn{2}{c}{Energy} & \multicolumn{2}{c}{Forces} \\
               Facet & RMSE & MAE & RMSE & MAE & RMSE & MAE & RMSE & MAE \\\hline
                & \multicolumn{8}{c}{Initial model} \\\hline
               (111) & 159.9 & 128.0 & 149.7 & 106.2 & 54.4 & 51.9 & 12.8 & 7.1 \\
               (100) & 262.9 & 165.1 & 254.7 & 94.6 & 60.6 & 57.5 & 17.7 & 8.7 \\
               (110) & 202.4 & 139.4 & 199.1 & 112.2 & 56.4 & 52.1 & 17.0 & 9.8 \\
               (211) & 334.8 & 251.2 & 329.2 & 148.9 & 62.5 & 56.0 & 13.8 & 9.2 \\ 
               \textbf{All} & \textbf{259.3} & \textbf{175.9} & \textbf{257.9} & \textbf{117.0} & \textbf{58.5} & \textbf{54.2} & \textbf{15.9} & \textbf{8.7} \\ \hline
                & \multicolumn{8}{c}{Final model} \\\hline
               (111) & 51.8 & 28.6 & 48.4 & 23.5 & 28.6 & 24.9 & 16.8 & 6.4 \\
               (100) & 73.5 & 41.6 & 89.7 & 30.9 & 27.8 & 24.9 & 12.9 & 6.8 \\
               (110) & 61.8 & 37.5 & 60.4 & 31.1 & 31.0 & 27.9 & 12.8 & 7.5 \\
               (211) & 72.7 & 51.0 & 58.6 & 36.1 & 29.9 & 26.2 & 11.1 & 7.1 \\ 
               \textbf{All} & \textbf{67.2} & \textbf{39.2} & \textbf{68.3} & \textbf{30.2} & \textbf{29.5} & \textbf{26.0} & \textbf{13.7} & \textbf{7.0} \\ \hline \hline
           \end{tabular}
           \label{tab:error_surfaces}
        \end{table}

        For further analysis, we have tabulated the RMSEs and MAEs of SchNet and PaiNN models averaged over 5 initial and 5 final SchNet and PaiNN models, differing only by training-validation-test split (Tab.~\ref{tab:error_surfaces}). Across all surface facets, the PaiNN model based on the initial data set provides energy predictions that are comparable to the SchNet model based on the final data set and force predictions that are vastly superior. Surprisingly, even when trained on a massively improved training data set, the PaiNN model force error is not drastically improved between the final and initial model. Of course, if adaptive sampling had been performed with the PaiNN model, it might have been possible to further reduce the force prediction error. Nevertheless, even for the final models, the average errors obtained with PaiNN are significantly lower than the lowest errors obtained with SchNet.
        
        Apart from the previous conclusions about the overall superiority of PaiNN over SchNet with respect to the errors, we can also deduce that the PaiNN models provide an improved prediction of sparsely sampled structures compared to SchNet models. For example, the H\textsubscript{2}/Cu(211) structures, for which no scattering data was included in our initial data set and thus a very limited number of data points are available (Fig.~\ref{fig:adaptive_sampl_h2oncu_xyz}.D and~S3), are predicted with high errors by the initial SchNet model. In contrast, the energy MAEs and RMSEs predicted by initial PaiNN models improve over the SchNet errors by more than 3 times, and the force errors improve by more than 13 times. Furthermore, both energy and force errors for H\textsubscript{2}/Cu(211) generated with PaiNN models are comparable to the errors obtained for the other surfaces, which is not the case for the SchNet models. This means that PaiNN not only improves greatly in terms of accuracy but also in terms of generalization across different systems (facets). As a result, it describes barriers across all facets very accurately.
        Comparing the errors obtained with initial and final models using SchNet, we notice the significant error improvement within all surfaces and the reduction of error prediction for H\textsubscript{2}/Cu(211) structures down to the average error level upon including adaptive sampling data for this surface. 
        
    \subsection{Scattering at high surface temperature} \label{sec:results_model_performance}

        Having constructed accurate and robust MLIPs for the prediction of sticking probabilities at 0~K that are comparable to previously published theoretical values, we finally investigate how general and robust the models are by comparing them against experimental sticking probabilities measured for different surface facets at high temperatures. Kaufmann~\textit{et~al.}~\cite{kaufmann_associative_2018} measured the sticking probabilities for H\textsubscript{2} on the Cu(111) and Cu(211) surface facets in the ground and excited rovibrational states. The H\textsubscript{2} ($\mathrm{\nu}$=1) sticking probabilities at both surfaces were measured at 923$\pm$3~K. This provides for an interesting benchmark of our models as we have not trained the models on high-temperature configurations of H\textsubscript{2} on Cu(111) or Cu(211), although we have included high-temperature displacements of the clean surface in the initial training data set.

        \begin{figure}
            \centering
            \includegraphics[width=2.3in]{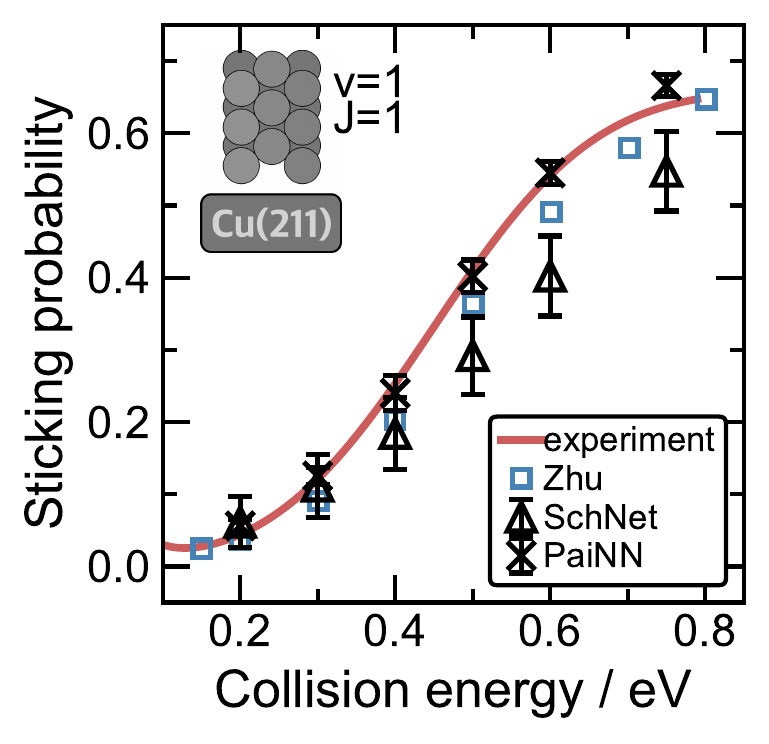}
            \caption{\textbf{SchNet and PaiNN sticking probabilities for H\textsubscript{2} scattering on Cu(211) at 925~K.} Probabilities were calculated at different collision energies using SchNet ($\pmb{\triangle}$) and PaiNN ($\pmb{\times}$) models for the ($\mathrm{\nu}$=1, J=1) rovibrational states. Each point represents an averaged value of sticking probability predicted using a committee of 3 models based on different train/test splits. For each initial condition, 10,000 trajectories were used for ensemble averaging. Blue squares (\textbf{{\color{MidnightBlue}$\pmb{\square}$}}) indicate the theoretical results obtained by Zhu~\textit{et~al.} (EANN model)~\cite{zhu_unified_2020}. The red line represents a sticking function obtained from the experimental results reported by Kaufmann~\textit{et~al.}~\cite{kaufmann_associative_2018}(923$\pm$3~K), scaled to match the theoretical probabilities obtained with the EANN model~\cite{zhu_unified_2020} at the highest collision energy (saturation parameter A=0.66 for both sticking functions).}
            \label{fig:sticking_v1j1_cu211}
        \end{figure}

        To confirm that our models can be employed at higher temperatures, we calculated sticking probabilities for H\textsubscript{2} at a 925~K Cu(211) surface in a rovibrationally excited state ($\mathrm{\nu}$=1, J=1). We compare our results to theoretical results calculated with the high-dimensional MLIP constructed by Zhu~\textit{et~al.}~\cite{zhu_unified_2020} and experimental results by Kaufmann~\textit{et~al.}~\cite{kaufmann_associative_2018} (see Fig.~\ref{fig:sticking_v1j1_cu211}). Note that the absolute sticking probabilities from permeation experiments cannot be directly compared to simulations.~\cite{kaufmann_associative_2018} A direct comparison to molecular beam scattering experiments \cite{rettner_quantumstatespecific_1995, michelsen_effect_1993} which provide absolute sticking probabilities may be preferable, but for the purpose of model assessment, the current comparison will suffice. To enable comparison to the permeation experiments, Zhu~\textit{et~al.} have scaled the experimental results of Kaufmann~\textit{et~al.} such that the probabilities of the experiment and calculation match for the highest reported collision energy of 0.8~eV. We similarly scale the sticking curve to the results of Zhu~\textit{et~al.}, which is shown in Fig.~\ref{fig:sticking_v1j1_cu211} (red line). We compare our independent simulation results to the literature data without any further scaling. Despite the fair agreement of SchNet with the reference data, the uncertainty of predictions made with the SchNet models is significantly higher than that obtained with the PaiNN models, which confirms the higher stability achieved by the PaiNN models. The sticking probabilities obtained with the SchNet models are noticeably lower than the rest of the models for collision energies above 0.4~eV. On the other hand, the results obtained with the PaiNN models better match the literature theoretical values and the experimental curve. This provides evidence that the models also perform well for high-temperature scattering where surface atom motion cannot be neglected. While scattering at 0~K surfaces can be approximated well with the BOSS approach in an effective six-dimensional PES, the high-temperature scattering requires dynamical sampling of surface degrees of freedom which is straightforward with the high-dimensional MLIPs we have constructed here, and the one used by Zhu~\textit{et~al.}~\cite{zhu_unified_2020}. 
        
        \begin{figure}
            \centering
            \includegraphics[width=3.4in]{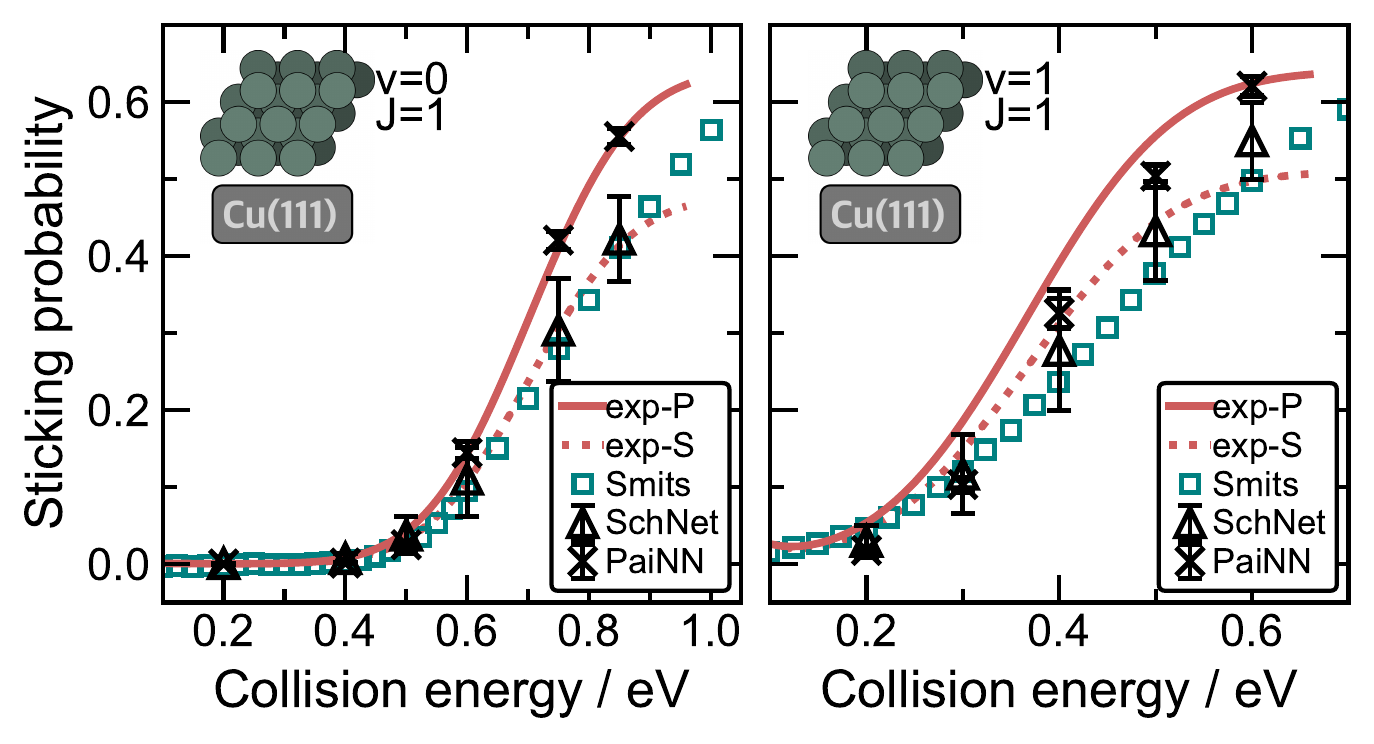}
            \caption{\textbf{SchNet and PaiNN sticking probabilities for H\textsubscript{2} scattering on Cu(111) at 925~K.} Probabilities were calculated at different collision energies using SchNet ($\pmb{\triangle}$) and PaiNN ($\pmb{\times}$) models for the ground ($\mathrm{\nu}$=0) (left) and excited ($\mathrm{\nu}$=1) (right) vibrational states (J=1 in both cases). Each point represents an averaged value of sticking probability predicted using a committee of 3 models based on different train/test splits. For each initial condition, 10,000 trajectories were used for ensemble averaging. Green squares (\textbf{{\color{teal}$\pmb{\square}$}}) indicate theoretical (QCD-EAM-DCM) results by Smits~\textit{et~al.}~\cite{smits_quantum_2022}. Red lines represent sticking functions obtained from the experimental results reported by Kaufmann~\textit{et~al.}~\cite{kaufmann_associative_2018}(923$\pm$3~K). The solid lines (labeled as \textit{exp-P})  are scaled to match the PaiNN sticking probabilities at the highest incident energy (saturation parameter A=0.64 for both sticking functions). The dotted lines (labeled as \textit{exp-S}) are scaled to match the QCD-EAM-DCM results (saturation parameter A=0.48 for ($\mathrm{\nu}$=0) and A=0.51 for ($\mathrm{\nu}$=1)).
            }
            \label{fig:sticking_cu111_exp}
        \end{figure}

        Sticking probabilities of H\textsubscript{2} on Cu(111) have also been measured by Kaufmann~\textit{et~al.} at 923$\pm$3~K. Within theoretical studies, the majority of references are based on the BOSS approximation, and effective phonon corrections, which become increasingly inaccurate at elevated surface temperatures.~\cite{dutta_effect_2021} However, a recent QCD-EAM-DCM model reported by Smits~\textit{et~al.}~\cite{smits_quantum_2022} accounted for surface degrees of freedom. To this date, this system has not been investigated at high temperatures, and at specific rovibrational states, with high-dimensional MLIPs.
        
        Fig.~\ref{fig:sticking_cu111_exp} reports the sticking probabilities predicted by the best final SchNet and PaiNN models on Cu(111) at a surface temperature of 925~K for two vibrational states ($\mathrm{\nu}$=0) and ($\mathrm{\nu}$=1). We  have scaled the experimental curves to match the highest collision energy of the PaiNN model (exp-P in Fig.~\ref{fig:sticking_cu111_exp}) and of the SchNet model (exp-S in Fig.~\ref{fig:sticking_cu111_exp}). We apply the same scaling (saturation parameter A=0.64 for PaiNN, and A=0.48 for SchNet models) for both vibrational states. In previous sections, we demonstrated that the PaiNN models display a better performance, e.g. by providing a smoother energy landscape, lower prediction errors, and better agreement with other references for sticking at Cu(211), as shown in Fig.~\ref{fig:sticking_v1j1_cu211}. Another hint at the better performance of PaiNN models is the fact that a single saturation parameter can be used to scale both rovibrational states, as shown in Fig.~\ref{fig:sticking_cu111_exp}. This  is not the case with SchNet models, for which using the same saturation parameter, adjusted for the ground vibrational state ($\mathrm{\nu}$=0), leads to experimental curves that do not match the probabilities for the higher vibrational state ($\mathrm{\nu}$=1). Additionally, the PaiNN results for H\textsubscript{2} ($\mathrm{\nu}$=1) sticking are in excellent agreement with the shape of the sticking curve obtained from the high-temperature experiments and the model uncertainty is very low. The SchNet models again underestimate the sticking probability for H\textsubscript{2} (both $\mathrm{\nu}$=0 and $\mathrm{\nu}$=1) scattering at high incidence energies, when compared to the PaiNN results or exp-P experimental curves. This is true even if we factor in the sizable uncertainty of the results as estimated from the standard deviation between the committee of three models for most instances. 
        In the case of PaiNN H\textsubscript{2} ($\mathrm{\nu}$=1) results, most of the predicted probabilities are slightly below the experimental results, especially for low collision energies, however, the predictions follow the shape of the experimental curve well. This is a promising result, considering that we did not apply a separate scaling for the H\textsubscript{2} ($\mathrm{\nu}$=1) experimental results. 
    
        Fig.~\ref{fig:sticking_cu111_exp} additionally reproduces sticking probabilities reported with the DCM model (QCD-EAM-DCM) by Smits~\textit{et~al.}~\cite{smits_quantum_2022}. These results show closer agreement with the SchNet rather than with the PaiNN predictions. However, we find this to be coincidental, due to the obvious weaknesses of the SchNet potential, such as the inability to create a smooth potential and significantly higher RMSEs and MAEs. Another likely origin of discrepancy between PaiNN and QCD-EAM-DCM models lies within the modeling of surface motion, since the predictions of sticking probabilities made with analytical QCD-BOSS model, and PaiNN at 0~K show excellent agreement (as shown in Fig.~\ref{fig:sticking_cu111_painn}). The surface motion in PaiNN is included just as any other degree of freedom, through the atomic NNs, giving a unified model for the movement of all atoms, whereas in the case of QCD-EAM-DCM model, the surface degrees of freedom are modeled by a potential based on embedded atom method (EAM), developed by Sheng~\textit{et~al.}~\cite{sheng_highly_2011}, which was fitted to the \textit{ab~initio} database and corrected with experimental results. Notably, the first-principles calculations used to fit this potential are based on a different functional. Furthermore, Sheng~\textit{et~al.}~\cite{sheng_highly_2011} exclusively included data at 0 and 300~K to construct the EAM potentials, which does not guarantee extrapolation to 900~K.~\cite{rassoulinejad-mousavi_interatomic_2018}

        We can further elucidate why adaptive sampling for low-temperature scattering enables us to describe high-temperature scattering by studying the ability of the two final models to predict the thermal lattice expansion of copper. Supplementary Figure~S6 shows the dependence of the energy with respect to the copper lattice constant. We see that PaiNN is in close agreement with DFT over a lattice expansion/contraction of $\pm 5\%$, despite the fact that we have only generated training data in fixed-size unit cells. The ability to describe the lattice expansion arises from the use of extended unit cells and the inclusion of high temperature MD data of the clean surfaces.
        
        We can conclude that the PaiNN MLIP is able to faithfully describe reactive scattering as a function of kinetic energy and molecular quantum state for a variety of surface facets and temperatures. Based on the same training data, the SchNet model was not able to achieve this to the same level, despite multiple rounds of adaptive sampling.

\section{\label{sec:conclusions}Discussion and Conclusions}

    We present a workflow to iteratively grow a training data set of DFT reference energies and forces based on adaptive sampling focused on improving the description of dynamic observables such as the sticking probability as a function of initial molecular quantum state and collision energy. We perform UQ based on ensemble learning throughout the adaptive sampling and calculate standard deviations on sticking probability predictions to indicate the epistemic error of the models. This is a generally applicable approach for the construction of MLIPs in gas-surface dynamics. 

    Utilizing this approach, we have compared the ability of two types of MLIPs based on MPNNs to describe the reactive scattering of molecules at metal surfaces, specifically the scattering and dissociative chemisorption of molecular hydrogen on multiple moving copper single-crystal surfaces. The two compared models are the invariant SchNet model and the equivariant PaiNN model. Both models are based on an MPNN with atom-centered descriptors that are trained against a combined energy and force loss function in an end-to-end fashion. The two models mainly differ in the nature of the descriptors: In the case of PaiNN, the atomic environment is also described with directional distance vectors, rather than just based on a spectrum of pairwise distances.
    
    Based on our analysis of the dynamical simulation of reactive surface chemistry, we can make the following observations:
    \begin{itemize}
    \item The SchNet model based on the initial data set was riddled with PES artifacts and even after several adaptive sampling iterations, the SchNet model was not able to provide a fully smooth energy landscape. While a significantly higher cutoff might potentially cure some of these artifacts, it would come at a significant increase in computational cost. In contrast, with the same cutoff, the equivariant MPNN  PaiNN was able to provide qualitatively correct and smooth PESs and (semi)quantitatively correct sticking probabilities using only the initial data set, reaching a fully converged quantitative description when including the adaptively sampled training data points.
    \item The PaiNN models, based on the same data, yield a beyond-five-fold reduction in force errors compared to the invariant MPNN SchNet and provide a better description of sparsely sampled PES regions. While SchNet models noticeably improve in later iterations with more evenly sampled training data, they remain less accurate than the PaiNN models. This finding is consistent with the benchmarks of PaiNN and NequIP against invariant architectures reported for the molecular dynamics simulations of small organic molecules (e.g. as represented by the MD17 data set~\cite{chmiela_machine_2017}).~\cite{schutt_equivariant_2021,batzner_e3-equivariant_2022}
    \item Adaptive sampling and active learning techniques have become a standard approach for the construction of robust and accurate MLIPs for molecular dynamics simulations of molecules and materials and have an important role to play to robustly predict dynamic observables in gas-surface dynamics and heterogeneous catalysis. Our results show that the equivariant PaiNN MLIPs generalize well based on sparse phase space data. This suggests that adaptive sampling with equivariant MPNNs can be performed based on a minimalistic initial data set to only generate as much reference data as absolutely necessary. This is a crucial benefit for the study of gas-surface dynamics where DFT reference data comes with a considerable computational cost. Additionally, our results show that UQ may be crucial in assessing the final success of every adaptive sampling iteration in predicting gas-surface-based reaction probabilities.
    \item The equivariant PaiNN model generalized better to different tasks and different conditions than SchNet. Our initial training data set was based on scattering simulations at frozen surfaces and AIMD simulations of clean single-crystal surfaces at different temperatures. All adaptive sampling was done to improve the description of scattering at 0~K surfaces. Yet, the PaiNN models were able to capture scattering at high-temperature surfaces, whereas the best SchNet model significantly underestimated sticking probabilities at high collision energies. Similarly, the PaiNN models have shown good performance in modeling scattering at the Cu(211) surface for our initial data set, in which structures sampled from scattering events at Cu(211) were lacking. Going beyond the cases covered by the current training data (e.g. dense hydrogen overlayers at copper surfaces) may require additional adaptive sampling steps.
    \end{itemize}

    The shortcomings of the SchNet model are based on its reliance on distance-only information within the cutoff region. Further details beyond the cutoff only enter indirectly via the message passing through the layers of the network. This appears to be insufficient to generate a smooth energy landscape of hydrogen-metal chemistry. Other successful (invariant) MLIPs, such as the EANN model or the  Behler-Parrinello NNs based on ACSFs, include 3-body terms within the descriptor which better resolve the local atomic environment, leading to smooth and accurate energy landscapes. Equivariant MPNNs such as PaiNN achieve the same by propagating vector-valued features through the network that can pass directional information between atoms. The latter has the benefit of reducing the number of hyperparameters associated with the initial basis definition, potentially providing a more user-friendly "black-box" approach. The equivariant PaiNN models proved their superiority over SchNet in every area of our study, however, we note that PaiNN requires more memory due to the inclusion of tensorial features, and thus for certain systems or computing architectures, it may be beneficial to employ rotationally invariant features.
    
    The excellent performance of equivariant MPNN-based MLIPs will, in the future, allow us to systematically build multi-purpose models for reactive chemistry. While adaptive sampling and ensemble learning allow us to improve model descriptions for specific dynamical observables (e.g. reactive scattering on single crystal surfaces), we can also assess and improve their ability to describe further related tasks (e.g. hydrogen evolution on crystalline nanoparticles) to systematically increase the range of applicability of an MLIP. Gas-surface dynamics puts an extreme demand on the accuracy and efficiency of MLIPs. Reaction probabilities change drastically with a subtle change of barriers and non-equilibrium events with low reaction probabilities require sampling of 10$^4$ to 10$^6$ trajectories to achieve meaningful statistical convergence. 
    While equivariant MPNN models appear to already satisfy the requirement of accuracy, their inference performance needs to further improve to advance the study of chemical dynamics at surfaces. Both SchNet and PaiNN models used in this study already improve roughly 10$^6$ times over DFT. Smaller NN models such as the EANN~\cite{zhang_embedded_2019} or linear models such as the atomic cluster expansion (ACE)~\cite{drautz_atomic_2019} offer high prediction efficiency, but may not offer as much generalization, which means that a trade-off might need to be sought between MLIPs that are general-purpose and MLIPs that have high evaluation efficiency.

\section*{ORCID iDs}
    Wojciech G. Stark \url{https://orcid.org/0000-0001-6279-2638}\\
    Julia Westermayr \url{https://orcid.org/0000-0002-6531-0742}\\
    Oscar A. Douglas-Gallardo \url{https://orcid.org/0000-0002-2250-3894}\\
    James Gardner \url{https://orcid.org/0000-0003-1840-804X}\\
    Scott Habershon \url{https://orcid.org/0000-0001-5932-6011}\\
    Reinhard J. Maurer \url{https://orcid.org/0000-0002-3004-785X}

\section*{Acknowledgments}
    This work was financially supported by The Leverhulme Trust (RPG-2019-078), the UKRI Future Leaders Fellowship programme (MR/S016023/1), and the Austrian Science Fund (FWF) [J 4522-N]. High-performance computing resources were provided via the Scientific Computing Research Technology Platform of the University of Warwick, the EPSRC-funded Materials Chemistry Consortium for the ARCHER2 UK National Supercomputing Service (EP/R029431/1), and the EPSRC-funded HPC Midlands+ computing centre (EP/P020232/1). We thank Bin Jiang (USTC, Hefei) for providing us with the training and test data set reported in Ref.~\cite{zhu_unified_2020}. We also thank Nils Hertl (Warwick), Daniel Auerbach (MPIBPC, Göttingen), and Sven Schwabe (IFNANO, Göttingen) for helpful discussions regarding the experimental results reported in Ref.~\cite{kaufmann_associative_2018}.

\section*{Data Availability}
The files containing inputs and outputs of the AIMD simulations and the DFT single-point calculations used in this study can be found within the NOMAD repository: \linebreak \href{https://nomad-lab.eu/prod/v1/gui/dataset/id/4r6KEYExRb-lQRUtpCfL7g}{10.17172/NOMAD/2023.05.03-2}.

\section*{Code Availability}
The molecular dynamics simulations were performed with the publicly available open-source code NQCDynamics.jl. The source code with detailed documentation, including many examples, is available on GitHub: \url{https://nqcd.github.io/NQCDynamics.jl/stable/}.
The dynamics for high-error structure search and the clustering scripts are available in the GitHub repository: \url{https://github.com/wgst/ml-gas-surface}. Detailed instructions for developing and using gas-surface MLIPs with the presented workflow can be accessed here: \url{https://wgst.github.io/ml-gas-surface/}.

\section*{Competing Interests}
The authors declare no competing financial or non-financial interests.

\section*{Author Contributions}
\textbf{Wojciech Stark:}
conceptualization (supporting);
data curation (lead);
investigation (lead);
software (lead);
validation (lead);
visualization (lead);
writing -- original draft (lead);
writing -- review and editing (equal).
\textbf{Julia Westermayr:} 
conceptualization (supporting);
investigation (supporting);
software (supporting);
writing -- review and editing (supporting).
\textbf{Oscar A. Douglas-Gallardo:}
investigation (supporting);
software (supporting).
writing -- review and editing (supporting).
\textbf{James Gardner:}
software (supporting);
writing -- review and editing (supporting).
\textbf{Scott Habershon:}
supervision (supporting);
writing -- review and editing (supporting).
\textbf{Reinhard J. Maurer:}
conceptualization (lead);
investigation (equal);
supervision (lead);
writing -- review and editing (equal). \\

\def\bibsection{\section*{\refname}} 
\bibliography{main}
\mbox{}
\newpage



\section*{Supplementary material}

# Supplementary Information to "Machine Learning Interatomic Potentials for Reactive Hydrogen Dynamics at Metal Surfaces Based on Iterative Refinement of Reaction Probabilities"

Wojciech G. Stark,[1] Julia Westermayr,[1, 2] Oscar A. Douglas-Gallardo,[1, 3]
James Gardner,[1] Scott Habershon,[1] and Reinhard J. Maurer[1, 4, *]

[1]*Department of Chemistry, University of Warwick,
Gibbet Hill Road, Coventry CV4 7AL, United Kingdom*
[2]*Present address: Wilhelm Ostwald Institute for Physical and
Theoretical Chemistry, University of Leipzig, Leipzig 04103, Germany*
[3]*Present address: Instituto de Ciencias Químicas, Facultad de Ciencias,
Universidad Austral de Chile, Isla Teja, Valdivia 5090000, Chile*
[4]*Department of Physics, University of Warwick,
Gibbet Hill Road, Coventry CV4 7AL, United Kingdom*

## I. STATISTICAL ERRORS EVALUATION

To calculate the statistical error of the sticking probability evaluation, based on 10,000 trajectories, we chose the following approach. First, we established 4 main groups, each containing 10,000 trajectories, and divided them into smaller subgroups, containing 500, 1,000, 2,000, and 2,500 trajectories (e.g. the first subgroup contains 20 subgroups of 500 trajectories). Next, we calculated sticking probabilities with the subgroups and the variance between them. Then we randomly shuffle the initial 10,000 trajectories and repeat the procedure 50 times. That allows us to obtain the linear fit of the inverse variance for the 10,000 trajectories (Fig. S1 (left)). Finally, the final error is obtained, by calculating the standard deviation. In Fig. S1 (right), we plotted the final sticking probabilities obtained using both SchNet and PaiNN models for $H_2$ ($\nu$=0, J=1) Cu(111) in 925 K with the error bars that include only statistical error, without the uncertainties included in the main manuscript. The obtained statistical errors are negligible, unlike the model uncertainties received from evaluating the predictions using the committee of 3 models, based on different train/test splits.

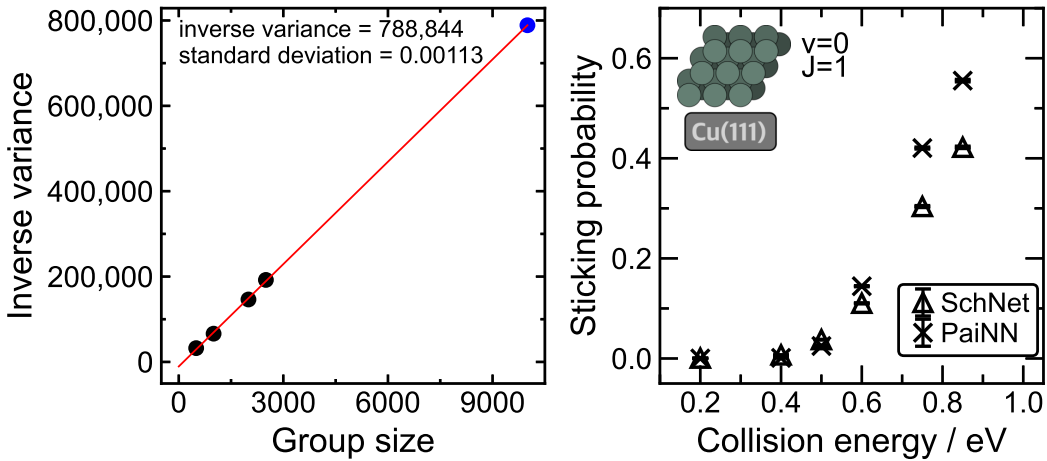

FIG. S1. **Statistical error evaluation.** The inverse variance of the final group size (10,000) predicted by linear regression fitted to four smaller group sizes (500, 1,000, 2,000, and 2,500) for collision energy of 0.6 (left side). Corresponding error bars for sticking probabilities predicted at different collision energies for $H_2$ ($\nu$=0, J=1) dissociative adsorption on Cu(111) in 925 K (right side).

---

[*] r.maurer@warwick.ac.uk

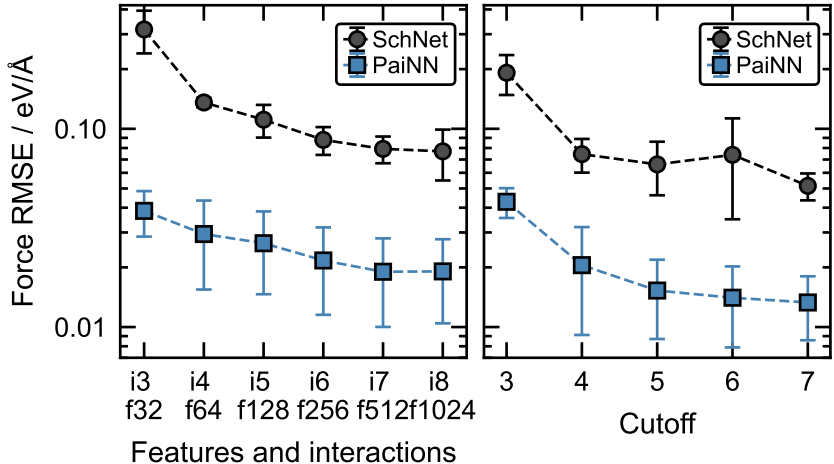

FIG. S2. **Optimization of model parameters.** The number of features and interaction blocks (left side) and the cutoff distance (right side) for non-equivariant (SchNet) and equivariant (PaiNN) MPNN models, based on the validation force RMSEs averaged over all of the cross-validation splits. The error bars represent the standard deviation between the force RMSEs obtained for all the splits. Optimization of the number of features and interaction blocks was done with the cutoff value of 4 Å and optimization of cutoff was done with 512 features and 7 interaction blocks.

## II. MODEL PARAMETER OPTIMIZATION

We performed an optimization of different model parameters, such as the number of features, the number of interaction blocks, and the cutoff distance. The optimization was performed for both SchNet and PaiNN models independently.

On the left side of Fig. S2 the convergence of the number of features and interaction blocks for both models with respect to force RMSEs in log-scale is presented. SchNet converges slightly quicker towards the optimal number of features and interactions, however, the force errors are considerably higher than what is found with PaiNN. Both models converge with respect to the force RMSEs with around 6-7 interaction blocks and 256-512 features. In order to obtain reliable and accurate models, despite the slower evaluation times, we have chosen to use 7 interactions and 512 features in the final models for both codes.

Overall, the convergence of features and interaction blocks for both models is relatively slow. The models require a large number of features to capture many-body interactions well and consequently to provide accurate energies and forces. That may be due to the high average coordination of the atoms in the surface slab systems and the different surface terminations included in the database.

The right side of Fig. S2 shows a plot of the force RMSE in the logarithmic scale against the cutoff distance, revealing that although PaiNN achieves smoother and quicker convergence, a satisfactory level of convergence can already be achieved with a cutoff distance of 4 Å with both methods. The quick convergence towards a relatively low cutoff distance that allows modeling $H_2$ molecule dynamics on Cu surfaces indicates that the interaction between hydrogen and the surface atoms is well captured by the message passing through 7 layers and not much is to be gained by increasing the cutoff beyond the 4 Å range. Extending the cutoff further would also come with a significant increase in computational cost to evaluate the model.

## III. DATABASE

The distribution of hydrogen atoms as a function of H-H and H-Cu distance is shown in Fig. S3. The figure shows that the distribution of H atoms in our database is very dense for the scattering simulations, all possible distances of hydrogen from the surfaces, as well as between the hydrogen atoms, are explored well.

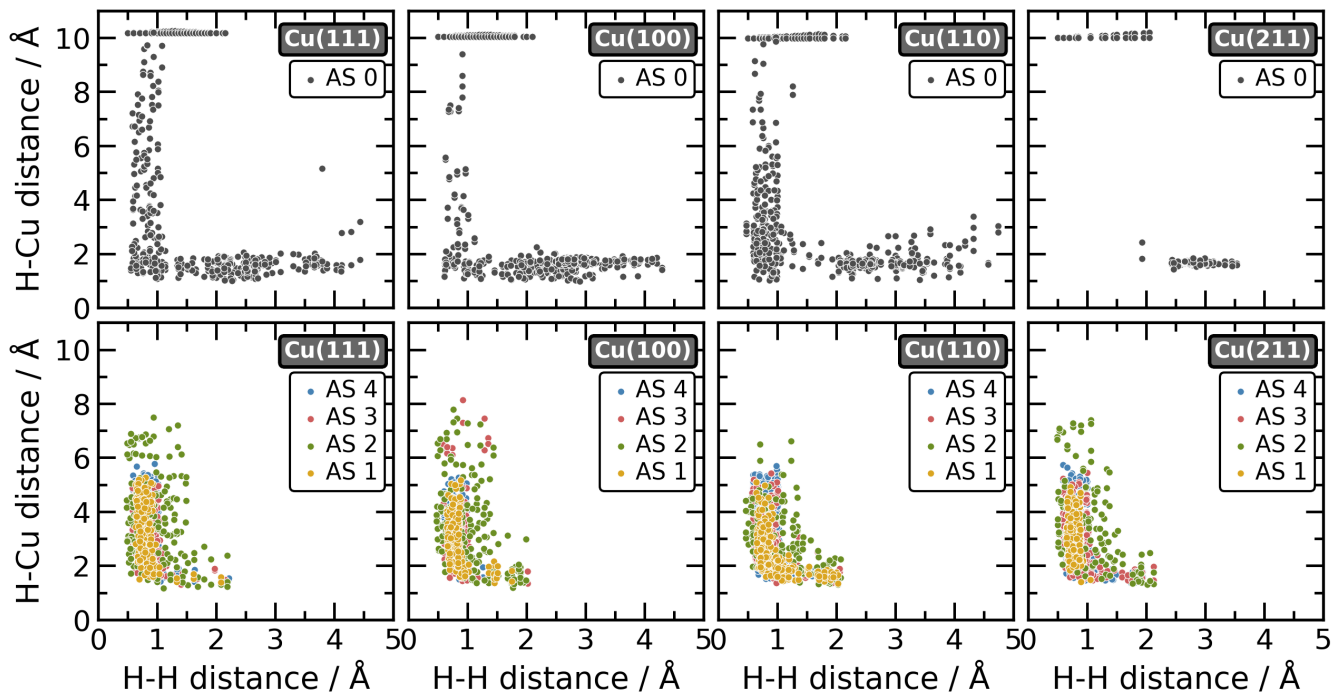

FIG. S3. **Distribution of hydrogen atoms depicted within H-H and H-Cu distances in our final database** The distribution includes datapoints from the initial database (*AS 0*) and following 4 iterations of adaptive sampling (*AS 1-4*). For calculations of H-Cu distance, we assume that Cu is the surface atom closest to the considered H atom. Note that the data points added through adaptive sampling (second row) are truncated at the H-H distance of 2.25 Å due to the stopping of dynamics trajectories.

## IV. MINIMUM ENERGY PATHS

Minimum energy paths (MEP) were obtained using climbing image nudged elastic bands (CI-NEB) employing DFT, SchNet, and PaiNN (Fig. S4). The MEPs obtained with DFT match the MEPs obtained with PaiNN models very well, suggesting that barriers are modeled remarkably well by PaiNN models. However, the MEPs generated with SchNet confirm the findings shown in Fig. 7, where the PES generated with SchNet is not smooth. The instabilities in PES obtained with SchNet cause inaccurate MEP predictions (especially for Cu(110)). The higher number of images (50 instead of 20) used with PaiNN code, as opposed to SchNet code, was possible due to the much faster convergence of the NEB paths (fewer NEB steps needed to converge).

## V. MODEL ERRORS

Fig. S5 shows energy and force predictions made with the final SchNet and PaiNN models compared to the reference DFT-based results. Energy RMSEs and MAEs correspond to the energy of an entire system, containing 56 atoms in $H_2$/Cu structures or 54 atoms in Cu surface-only structures.

## VI. DEPENDENCE OF ENERGY WITH RESPECT TO THE COPPER LATTICE CONSTANT

The potential energy values obtained for specific lattice constant parameters calculated with DFT, SchNet, and PaiNN are shown in Fig. S6. We can clearly observe that energy predictions for different lattice constants are excellent. In contrast, the SchNet clearly struggles with the predictions, especially outside of the close range around the equilibrium lattice constant.

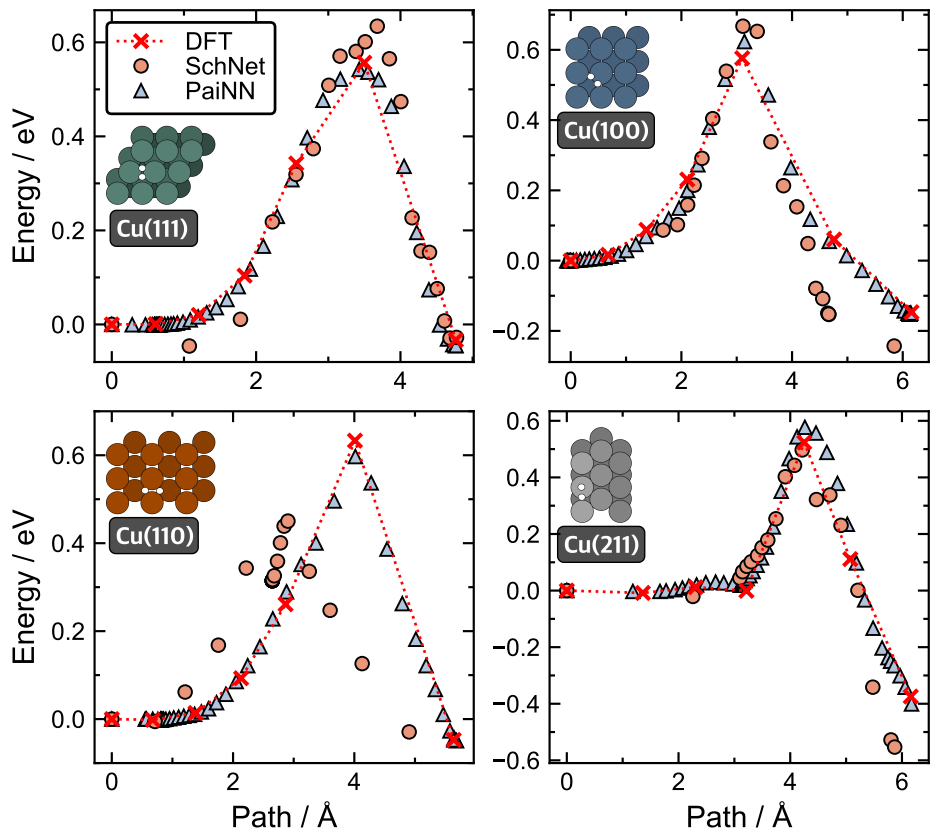

FIG. S4. **Minimum energy paths obtained using CI-NEB method for $H_2$ dissociative adsorption on different copper surfaces.** Potential energy is shown along the reaction path (Å), calculated using DFT (×) and both of the MLIPs included in our study: SchNet (orange circles) and PaiNN (grey-blue triangles). Figures that show top-down view of the systems at the transition states are included on the respective plots.

S4

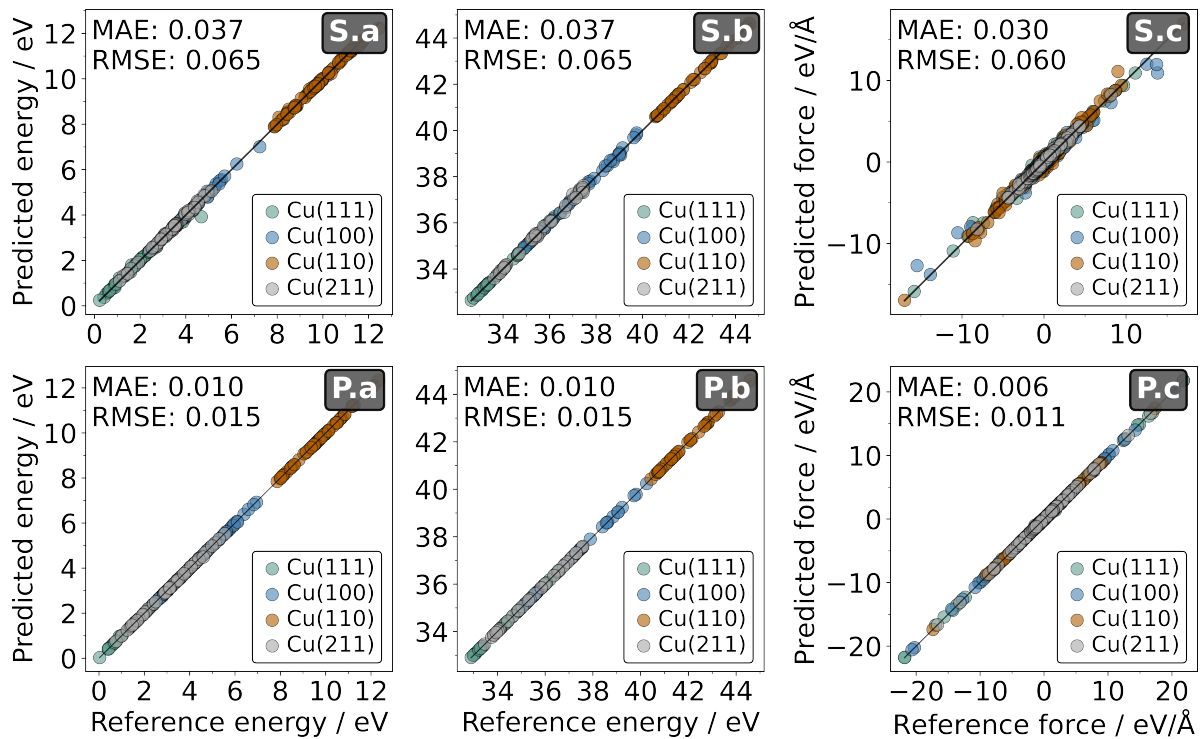

FIG. S5. **Energy (*a,b*) and force (*c*) predictions for $H_2$/Cu structures using SchNet (*S.a-c*) and PaiNN (*P.a-c*).** Best final models are compared to the reference (DFT) energies and forces. *S.a* and *P.a* show energies predicted for the structures with both $H_2$ and Cu atoms, *S.b* and *P.b* show energies predicted for just Cu surface structures. Force predictions include all of the structures together. The values of corresponding MAEs and RMSEs for energy (eV) or force (eV/Å) predictions are displayed in the upper left corner of every plot.

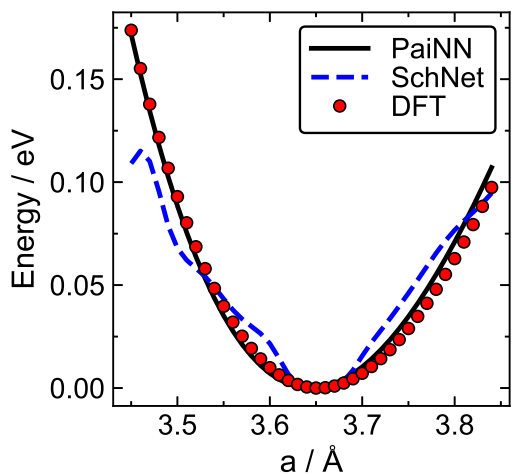

FIG. S6. **Relative potential energies of the Cu atoms for a set of lattice constants a (Å).** Energies were calculated using a DFT code (red data points) and two MLIPs included in our study, namely, SchNet (blue, dashed line) and PaiNN (black, solid line).



\end{document}